%% file: fouco.tex
\documentclass[preprint,draft]
{sigplanconf}

\usepackage[
   a4paper,
   pdftex,
   pdfkeywords={},
   pdfborder={0 0 0},
   pdftitle={Foundational Extensible Corecursion},
   pdfauthor={Jasmin Christian Blanchette, Andrei Popescu, and Dmitriy Traytel},
   draft=false,
   bookmarksnumbered,
   bookmarks,
   bookmarksdepth=2,
   bookmarksopenlevel=2,
   bookmarksopen]{hyperref}

\usepackage{flushend}
\usepackage{quoting}
\usepackage[T1]{fontenc}
\usepackage[all]{xy}
\usepackage{amsmath}
\usepackage{amsthm}
\usepackage{wasysym}
\usepackage{bm}
\usepackage{bussproofs}
\usepackage{caption}
\usepackage{subcaption}
\usepackage{mathptmx}
\usepackage{multicol}
\usepackage{url}
\usepackage[scaled=.825]{beramono}
\usepackage[scaled=.825]{helvet}
\usepackage{tikz}
\usetikzlibrary{positioning,shapes,arrows,matrix}

\include{defs}

\makeatletter

\DeclareSymbolFont{letters}{OML}{txmi}{m}{it} 
\EnableBpAbbreviations

\def\S{Section~}

\begin{document}

\balancecolumns



\pagestyle{plain} 

\title{Foundational Extensible Corecursion}

\authorinfo{Jasmin Christian Blanchette}
{Inria Nancy \& LORIA, France \\ Max-Planck-Institut f\"ur Informatik, Saarbr\"ucken, Germany}
{\texttt{jablanch@inria.fr}}

\authorinfo{Andrei Popescu}
{Department of Computer Science,  \\
School of Science and Technology, \\
Middlesex University, UK
}
{\texttt{a.popescu@mdx.ac.uk}}

\authorinfo{Dmitriy Traytel}
{Technische Universit\"at M\"{u}nchen, Germany}
{\texttt{traytel@in.tum.de}}


\maketitle

\newcommand\XXX{\textbullet}
\newbox\boxHyph\setbox\boxHyph=\hbox{\textbf{-}}
\newbox\boxX\setbox\boxX=\hbox{\XXX}

\renewcommand\UrlFont{\tt}

\begin{abstract}
This paper presents a formalized framework for defining corecursive functions
safely in a total setting, based on corecursion up-to and
relational parametricity. The end product is a general corecursor that allows
corecursive (and even recursive) calls under well-behaved operations, including
constructors. Corecursive functions that are well behaved can be registered as
such, thereby increasing the corecursor's expressiveness.
The metatheory
is formalized in the Isabelle proof assistant and forms the core of a prototype
tool. 
The corecursor is derived from first
principles, without requiring new axioms or extensions of the logic.
\end{abstract}



{\raggedright 
 \\
higher-order logic,
Isabelle
}

\section{Introduction}
\label{sec-intro}

Total functional programming is a discipline that ensures computations
always terminate. It is invaluable in a proof assistant, where nonterminating
definitions such as $\const{f}\;x = \const{f}\;x + 1$ can yield
contradictions. Hence, most 
assistants will accept recursive functions
only if they can be shown to terminate. Similar concerns arise in specification
languages 
and verifying compilers.

However, some processes need to run forever, without their being inconsistent.
An important class of total programs has been identified under the heading of
\emph{productive coprogramming}
\cite{turner-1995,abbott-et-al-2005,mcbride-productive}: These
are functions that progressively reveal parts of their (potentially infinite)
output. For example, given a type of infinite streams constructed by $\SCons$,
the definition
\[\const{natsFrom}\;n  =  \SCons\;n\;(\const{natsFrom}\;(n + 1))\]
falls within this fragment, since each call to \const{natsFrom} produces one
constructor before entering the nested call. Not only is the equation
consistent, it also fully specifies the function's behavior.

The above definition is legitimate only if objects are allowed to be infinite.
This may be self-evident in a nonstrict functional language such as Haskell, but
in a total setting we must carefully distinguish between the well-founded
inductive (or algebraic) datatypes and the non-well-founded
coinductive (or coalgebraic) datatypes---often simply called
\emph{datatypes} and \emph{codatatypes}, respectively.
\emph{Recursive} functions consume datatype values, peeling off a finite number
of constructors as they proceed; \emph{co\-recursive} functions produce
codatatype values, consisting of finitely or infinitely many constructors. And
in the same
way that \emph{induction} is available as a proof principle to reason about
datatypes and terminating recursive functions, \emph{coinduction}
supports reasoning over codatatypes and productive corecursive functions.

Despite their reputation for esotericism, codatatypes have an
important role to play in both the theory and the metatheory of programming. On
the theory side, they allow a direct 
embedding of a large class of
nonstrict functional programs in a total logic. In conjunction
with interactive proofs and code generators, 
this enables certified functional programming \cite{benton-integrated}.
On the metatheory side, codatatypes conveniently capture infinite, possibly
branching processes. Major proof developments rely on them, including those
associated with a C compiler \cite{leroy-2009}, a Java compiler
\cite{lochbihler-2010-jinja}, and the Java memory model
\cite{lochbihler-2014-jmm}.
\leftOut{Beyond
programming, a number of papers illustrate how coinductive methods can lead to
more elegant solutions than traditional approaches. A recent instance is a formal proof
of the completeness theorem for classical first-order logic
\cite{blanchette-et-al-2014-compl}, 
failed attempts at finding a proof give rise to
based on possibly infinite derivation trees.}

\newcommand\NOCITE[1]{}

Codatatypes are supported by an increasing number of proof assistants, including
Agda\NOCITE{bove-et-al-2009}, Coq\NOCITE{bertot-casteran-2004},
Isabelle\slash HOL\NOCITE{nipkow-et-al-2002},
Isabelle\slash ZF\NOCITE{paulson-1993-zf,paulson-1995-zf}, Matita\NOCITE{asperti-et-al-2011},
and PVS\NOCITE{crow-et-al-1995}. They are also present in
the CoALP dialect of logic programming\NOCITE{heras-et-al-201x} and in
the Dafny specification language\NOCITE{leino-moskal-2014}.
But the ability to introduce codatatypes is not worth much without adequate support
for defining meaningful functions that operate on them. For most systems, this support
can be characterized as work in progress. The key question they all must answer is:
\emph{What right-hand sides can be safely allowed in a function definition?}

Generally, there are two main approaches to support recursive and corecursive
functions in a proof assistant or similar system:
\begin{description}
\item[The intrinsic approach:]
A syntactic criterion is built into the logic:\ termination for recursive
specifications, productivity (or guardedness) for corecursive specifications.
The termination or productivity checker is part of the system's trusted code
base.

\item[The foundational approach:]
The (co)recursive specifications are reduced to a fixpoint construction,
which permits a simple definition of the form $\const{f} = \ldots$\,,
where \const{f} does not occur in the right-hand side.
The original equations are derived as theorems from this internal definition,
using dedicated proof tactics.
\end{description}
Systems favoring the intrinsic approach
include the 
proof assistants Agda and Coq, as well
as tools such as CoALP and Dafny. The main hurdle for their
users is that syntactic criteria are inflexible; the specification must be
massaged so that it falls within a given
syntactic fragment, even though the desired
property (termination or productivity) is semantic. But perhaps more troubling in
systems that process theorems, soundness is not obvious at all and very
tedious to ensure; as a result, there is a history of critical bugs in
termination and productivity checkers, as we will see when we review related work
(\S\ref{sec-rel}). Indeed, Abel \cite{abel-2013-coqml} observed that
\begin{quote}
Maybe the time is ripe to switch to a more semantical notion of termination and
guardedness. The syntactic guard condition gets you somewhere, but then needs a
lot of extensions and patching to work satisfactory in practice. Formal
verification of it becomes too difficult, and only intuitive justification is
prone to errors.
\end{quote}

In contrast to Agda and Coq, proof assistants based on higher-order logic (HOL),
such as HOL4, HOL Light, and Isabelle\slash HOL, generally adhere to the
foundational approach. Their logic
is expressive enough to accommodate the (co)algebraic constructions underlying
(co)datatypes and (co)recursive functions in terms
of functors on the category of sets \cite{traytel-et-al-2012}.
The main drawback of this approach is that
it requires a lot of work, both conceptual and implementational. Moreover, it
is not available for all systems, since it requires an expressive enough logic.

Because every step must be formally justified, foundational definitional
principles tend to be simpler and more restrictive than their intrinsic
counterparts. As a telling example, codatatypes were introduced in
Isabelle\slash HOL only recently, almost two decades after their inception in
Coq, and they are still missing from the other HOL systems; and corecursion is
limited to the primitive case, in which corecursive calls occur under exactly
one constructor.

That primitive corecursion (or the slightly extended version supported by Coq)
is too restrictive is an observation that has been made repeatedly by
researchers who use corecursion in Coq and now also Isabelle.
Lochbihler and H\"olzl dedicated a paper \cite{lochbihler-hoelzl-2014}
to ad hoc techniques for defining operations on corecursive lists in Isabelle.
Only after introducing a lot of machinery do they manage to
define their central example---\const{lfilter}, a filter function
on lazy (coinductive) lists---and derive suitable reasoning principles.

We contend that it is possible to combine advanced features as found in
Agda and Coq with the fundamentalism of Isabelle.
The lack of built-in support for corecursion, an apparent weakness, reveals
itself as a strength as we proceed to introduce rich notions of corecursion,
without extending the type system or adding axioms.

In this paper, we formalize a highly expressive corecursion framework that extends primitive
corecursion in the following ways:
%
It allows corecursive calls under several constructors;
it allows well-behaved operators in the context around or
between the constructors and around the corecursive calls;
importantly, it supports blending terminating recursive calls 
with guarded corecursive calls.
%
This general corecursor is accompanied by a corresponding, equally general
coinduction principle that makes reasoning about it convenient. Each of the
corecursor, mixed recursor--corecursor, and the coinduction principle
{\relax grow in expressiveness
during the interaction with the user}, by learning of new well-behaved contexts.
The 
constructions draw heavily from category theory.

Before presenting the technical details, we first show through examples how a
primitive corecursor can be incrementally enriched to accept ever richer
notions of corecursive call context (\S\ref{sec-exa}). This is made possible by
the modular bookkeeping of additional structure for the involved type constructors,
including a relator structure. This structure can be exploited
to prove 
parametricity theorems, which 
ensure the suitability of operators as participants to the call contexts, in the style
of coinduction up-to. Each new corecursive definition is a
potential future participant (\S\ref{sec-meta}).

This extensible corecursor gracefully
handles codatatypes with nesting through arbitrary type constructors (e.g., for
infinite-depth Rose trees nested through finite or infinite lists).
Thanks to the framework's modularity,
function specifications can combine corecursion with recursion, 
yielding quite expressive mixed fixpoint 
definitions (\S\ref{sec-mixed}).
This is inspired by the Dafny tool, but our approach is
semantically founded and hence provably consistent.



The complete metatheory is implemented in Isabelle\slash HOL, as a combination of a generic
proof development parameterized by arbitrary type constructors and a tool for instantiating
the metatheory to user-specified instances (\S\ref{sec-for}, \cite{our-formalization}).

Techniques such as corecursion and coinduction up-to have been
known for years in the process algebra community, before they were embraced and
perfected by category theorists (\S\ref{sec-rel}).
This work is part of a wider program aiming at bringing
insight from category theory into proof assistants
\cite{traytel-et-al-2012,blanchette-et-al-2014-impl}. 
The main contributions of this paper are the following:
\begin{itemize}
\item We represent in higher-order logic an integrated framework for recursion and corecursion
able to evolve by user interaction.



\item We identify a sound fragment of mixed recursive--corecursive
specifications, integrate it in our framework, and present several examples
that motivate this feature.

\item We implement the above in Isabelle/HOL within an interactive loop that maintains the recursive--corecursive infrastructure.

\item We use this infrastructure to automatically derive many examples that are
problematic in other proof assistants.
\end{itemize}

A distinguishing feature of our framework is that it does not require the user
to provide type annotations. On the design space, it lies
between the highly restrictive primitive corecursion and
the more bureaucratic up-to approaches such as clock variables
\cite{mcbride-productive,clouston-et-al-2015}
and sized types \cite{abel-2004}, 
combining expressiveness and ease of use.
The identification of this ``sweet spot'' can also be seen as a contribution.

\section{Motivating Examples}
\label{sec-exa}




We demonstrate the expressiveness of the corecursor framework by
examples, adopting the user's perspective. The case studies by
Rutten~\cite{rutten05} and Hinze~\cite{hinze10} on stream calculi serve as
our starting point. Streams of natural numbers can be defined as
\begin{quote}
  \keyw{codatatype} \,$\Stream$ $=$ $\SCons\;(\hd: \Nat)\;(\tl: \Stream)$
\end{quote}
where $\SCons : \Nat
\ra \Stream \ra \Stream$ is the constructor and $\hd : \Stream \ra \Nat$, $\tl :
\Stream \ra \Stream$ are its selectors.
Although the examples may seem simple or contrived, they were carefully
chosen to show the main difficulties that arise in practice.

\subsection{Corecursion Up-to}
\label{sec-uptoExa}

As our first example of a corecursive function definition,
we consider the pointwise sum of two streams:
\begin{quote}
  $\xs \opls \ys  =  \GUARD{\SCons}\;(\hd\;\xs + \hd\;\ys)\;(\tl\;\xs \opls
  \tl\;\ys)$
\end{quote}
The specification is productive, since the corecursive call occurs directly
under the stream constructor, which acts as a guard (shown underlined).
Moreover, it is primitively corecursive, because the topmost symbol on the
right-hand side is a constructor and the corecursive call
appears directly as an argument to it.

These syntactic restrictions can be relaxed to allow conditional statements
and `let' expressions \cite{blanchette-et-al-2014-impl}, but despite
such tricks primitive corecursion remains hopelessly primitive.
The syntactic restriction for admissible corecursive definitions in Coq is more
permissive in that it allows for an arbitrary number of constructors to guard
the corecursive calls, as in the following 
definition:
\begin{quote}
$\onetwos  =  \GUARD{\SCons}\;1\;(\GUARD{\SCons}\;2\;\onetwos)$
\end{quote}

Our framework achieves the same result by registering $\SCons$ as a
well-behaved operation. Intuitively, an operation is \emph{well behaved} if it
needs to destruct at most one constructor of input to produce one
constructor of output.
For streams, such an operation may inspect the head and the tail
(but not the tail's tail) of its arguments before producing an $\SCons$.
Because the operation preserves productivity, it can safely
surround the guarding constructor.

The rigorous definition of well-behavedness 
will capture
this intuition in a parametricity property that must be discharged
by the user. In exchange, the framework yields
a strengthened corecursor that incorporates the new operation.

The constructor $\SCons$ is well behaved, since it does not even need to
inspect its arguments to produce a constructor.
In contrast, the selector $\tl$ is not well behaved---it must destruct two
layers of constructors to produce one:
\begin{quote}
$\tl\;\xs = \SCons\;(\hd\;(\tl\;\xs))\;(\tl\;(\tl\;\xs))$
\end{quote}
The presence of non-well-behaved operations in the corecursive call context
is enough to break productivity, as
in the example $\const{stallA}  =  \SCons\;1\;(\tl\;\const{stallA})$, which stalls
immediately after producing one constructor, leaving $\tl\;\const{stallA}$ unspecified.

Another instructive example is the function that keeps every other element in a stream:
\begin{quote}
$\everyOther\;\xs  =  \SCons\;(\hd\;\xs)\;(\everyOther\;(\tl\;(\tl\;\xs)))$\kern200mm 
\end{quote}
The function is 
not well behaved, despite being primitive corecursive.
It also breaks productivity: $\const{stallB}  =  \SCons\;1\;(\everyOther\allowbreak\;\const{stallB})$
stalls after producing two constructors.

Going back to our first example, we observe that the operation~$\opls$ is well
behaved. Hence, it is allowed to participate in corecursive call contexts when
defining new functions. In this respect, the framework is more permissive than
Coq's syntactic restriction. For example, we can define the stream of Fibonacci
numbers in either of the following two ways:
\begin{quote}
$\fibA  =  \GUARD{\SCons}\;0\;(\SCons\;1\;\fibA \vvthinspace\opls\vvthinspace \fibA)$ \\[1\jot]
$\fibB  =  \GUARD{\SCons}\;0\;(\SCons\;1\;\fibB) \vthinspace\opls\vthinspace \GUARD{\SCons}\;0\;\fibB$
\end{quote}
Well-behaved operations are allowed to appear both under the constructor guard
(as in $\fibA$) and around it (as in $\fibB$). Notice that two guards are
necessary in the second example---one for each branch of the $\opls$ operator.
Incidentally, we are not aware of any other framework that allows such definitions. Without
rephrasing the specification, $\fibB$ cannot be expressed in Rutten's format of
behavioral differential equations~\cite{rutten05} or in Hinze's syntactic
restriction~\cite{hinze10}, nor via Agda copatterns \cite{abelPTS-2013,abelP-2013}.

Many useful operations are well behaved and can therefore participate in further
definitions.
Following Rutten, 
the shuffle product $\oprd$ of
two streams is defined in terms of 
$\opls$. Shuffle product
being itself well behaved, we can employ it to define stream exponentiation,
which also turns out to be well behaved:
\begin{quote}
$\xs \oprd \ys  =  \GUARD{\SCons}\;(\hd\;\xs \times \hd\;\ys)\;\\
 \phantom{\xs \oprd \ys  = \GUARD{\SCons}\;}((\xs \oprd \tl\;\ys) \opls (\tl\;\xs \oprd ys))$\\[1\jot]
$\oexp\;\xs  =  \GUARD{\SCons}\;(2 \mathbin{\vthinspace\char`\^\vthinspace} {\hd\;\xs})\;(\tl\;\xs \oprd \oexp\;\xs)$
\end{quote}
%
Next, we use the defined and registered operations to specify two streams of
factorials of natural numbers $\facA$ (starting at 1) and $\facB$ (starting
at 0):%
\begin{quote}
$\facA  =  \GUARD{\SCons}\;1\;\facA \oprd \GUARD{\SCons}\;1\;\facA$\\[1\jot]
$\facB  =  \oexp\;(\GUARD{\SCons}\;0\;\facB)$
\end{quote}
Computing the first few terms of $\facA$ manually should convince the reader
that productivity and efficiency are not synonymous.

The arguments of well-behaved operations are not restricted to the $\Stream$
type. For example, we can define the
well-behaved supremum of a finite set of streams by primitive corecursion:
\begin{quote}
$\SUP\;X= \GUARD{\SCons}\;({\bigsqcup}\,(\fimage\;\hd\;X))\;(\SUP\;(\fimage\;\tl\;X))$
\end{quote}
Here, $\fimage$ gives the image of a finite set under a function,
and $\bigsqcup X$ is the maximum of a finite set of naturals or $0$ if
$X$ is empty.


\subsection{Nested Corecursion Up-to}
\label{sec-nest-exa}

Although we use streams as our main example, the framework generally supports
arbitrary codatatypes with multiple curried constructors and nesting through
other type constructors. To demonstrate this last feature, we introduce the type of
finitely branching Rose trees of potentially infinite depth with numeric labels:
\begin{quote}
  \keyw{codatatype} \,$\Tree$ $=$ $\Node\;(\Label: \Nat)\;(\Children: \List\;\Tree)$
\end{quote}
The type $\Tree$ has a single constructor $\Node : \Nat \ra \List\;\Tree \ra
\Tree$ and two selectors $\Label : \Tree \ra \Nat$ and $\Children : \Tree \ra
\List\;\Tree$. The recursive occurrence of $\Tree$ is {nested} in the
familiar polymorphic datatype of finite lists.

We first define the pointwise sum of two trees analogously to $\opls$:
\begin{quote}
  $t \mathrel\boxplus u  =  \GUARD{\Node}\;(\Label\;t + \Label\;u)\;\\
\phantom{t \mathrel\boxplus u  = \GUARD{\Node}\;}(\map\;(\lambda (t', u').~t' \mathrel\boxplus u')\;(\zip\;(\Children\;t)\;(\Children\;u)))\kern-200mm$
\end{quote}
Here, $\map$ is the standard map function on lists,
and $\zip$ converts two parallel lists
into a list of pairs, truncating the longer list if necessary.
The criterion for primitive corecursion for nested codatatypes requires
the corecursive call to be applied through $\map$, which is
the case for $\boxplus$. Moreover, by virtue of being well behaved,
$\boxplus$ can be used to define the shuffle product of trees:
\begin{quote}
$t \mathrel\boxtimes u  =  \GUARD{\Node}\;(\Label\;\xs \times \Label\;\ys)\;\\
 \phantom{t \mathrel\boxtimes u}(\map\;(\lambda (t', u').~(t \mathrel\boxtimes u') \mathrel\boxplus (t' \mathrel\boxtimes u))\;(\zip\;(\Children\;t)\;(\Children\;u)))\kern-200mm$
\end{quote}
Again, the corecursive call takes place inside $\map$, but this
time also in the context of $\boxplus$. The specification of
$\boxtimes$ is 
corecursive up-to and well behaved.

\subsection{Mixed Recursion--Corecursion}
\label{sec-mix-exa}

It is often convenient to let a corecursive function perform some finite
computation before producing a constructor. With mixed recursion--corecursion, a
finite number of unguarded recursive calls perform this calculation before
reaching a guarded corecursive call.

The intuitive criterion for accepting such definitions is that the unguarded
recursive call could be unfolded to arbitrary finite depth,
ultimately yielding a purely corecursive definition. An example is
the $\primes$ function taken from Di~Gianantonio and Miculan~\cite{miculan-unifying}:
\begin{quote}
$\primes\;m\;n  =  \keyw{if}~(m = 0 \mathrel\land n > 1) \mathrel\lor \GCD\;m\;n = 1\\
\phantom{\primes\;m\;n  =  {}} \keyw{then}~\GUARD{\SCons}\;n\;(\primes\;(m \times n)\;(n + 1))\\
\phantom{\primes\;m\;n  =  {}} \keyw{else}~\primes\;m\;(n + 1)$
\end{quote}
When called with $m = 1$ and $n = 2$, this function computes the stream of prime
numbers. The unguarded call in the $\keyw{else}$ branch increments its second
argument $n$
until it is coprime to the first argument~$m$ (i.e., the greatest common divisor
of $m$ and $n$ is $1$). For any positive integers $m$ and $n$,
the numbers $m$ and $m \times n + 1$ are
coprime, yielding an upper bound on the number of times $n$ is increased. Hence,
the function will take the \keyw{else} branch at most finitely often before
taking the \keyw{then} branch and producing one constructor. There is a slight
complication when $m = 0$ and $n > 1$: Without the first disjunct
in the $\keyw{if}$ condition, the function could stall. (This corner case
was overlooked in the original example~\cite{miculan-unifying}.)

Mixed recursion--corecursion also allows us to give a definition of factorials
without involving any auxiliary stream operations: 
\begin{quote}
$\facC\;n\;a\;i  =  \keyw{if}~i = 0~\keyw{then}~\GUARD{\SCons}\;a\;(\facC\;(n + 1)\;1\;(n + 1))\\
\phantom{\facC\;n\;a\;i  =  {}} \keyw{else}~\facC\;n\;(a \times i)\;(i - 1)$
\end{quote}
The recursion in the \keyw{else} branch computes the next
factorial by means of an accumulator $a$ and a decreasing counter $i$. When the
counter reaches $0$, $\facC$ corecursively produces a constructor with the
accumulated value and resets the accumulator and the counter.

Unguarded calls may also occur under well-behaved operations:
\begin{quote}
$\catalan\;n  =  \keyw{if}~n > 0~\keyw{then}~\catalan\;(n - 1) \opls \GUARD{\SCons}\;0\;(\catalan\;(n + 1))\kern-200mm\\
\phantom{\catalan\;n  =  {}} \keyw{else}~\GUARD{\SCons}\;1\;(\catalan\;1)$
\end{quote}
The call $\catalan\;1$ computes the stream of Catalan numbers:
$C_1,\allowbreak C_2, \ldots$\,, where $C_i = {1 \over n + 1}{2n \choose n}$.
This fact is far from obvious. Productivity is not entirely obvious either, but
it is guaranteed by the framework.

When mixing recursion and corecursion, it is easy to
get things wrong in the absence of solid foundations.
Consider this apparently unobjectionable specification in which the
corecursive call is guarded by $\SCons$ and the unguarded call's argument
strictly decreases toward 0:
\begin{quote}
$\evil\;n  =  \keyw{if}~n < 2~\keyw{then}~\SCons\;n\;(\evil\;(n+1))\\
\phantom{\evil\;n  =  {}} \keyw{else}~\inc\;(\tl\;(\evil\;(n-1)))$
\end{quote}
Here, $\inc = \mapS\;(\lambda x.~x + 1)$ and $\mapS$ is the map function on
streams. A simple calculation reveals
that this specification is inconsistent
because the $\tl$ selector before the unguarded
call destructs the freshly produced constructor from the other branch:
\begin{quote}
$\evil\;2{} = {}\inc\;(\tl\;(\evil\;1))$ \\
$\phantom{\evil\;2}{} = \inc\;(\tl\;(\SCons\;1\;(\evil\;2)))$\kern-200mm \\
$\phantom{\evil\;2}{} = {}\inc\;(\evil\;2)$
\end{quote}
This is a close cousin of the infamous $\const{f}\;x = \const{f}\;x + 1$ example
mentioned in the introduction.
The framework rejects this specification on the grounds that the $\tl$ selector in the
recursive call context is not well behaved.

We conclude this section with a practical example from the literature.
Given the polymorphic type
\begin{quote}
  \keyw{codatatype} \,$\TC{LList}(A)$ $=$ \\
  \noindent\hbox{}\quad $\const{LNil}$ $\mid$ $\const{LCons}\;(\hd: A)\;(\tl: \TC{LList}(A))$
\end{quote}
of lazy lists, the task is to define the function
$\const{lfilter} : (A \ra \Bool) \ra \TC{LList}(A) \ra \TC{LList}(A)$
that filters out all elements failing to satisfy the given predicate.
Thanks to the support for mixed recursion--corecursion, the framework
transforms what was for Lochbihler and H\"olzl \cite{lochbihler-hoelzl-2014} a
research problem into a routine exercise:
\begin{quote}
$\const{lfilter}\;P\;\mathit{xs} =
      \!\begin{aligned}[t]
& \keyw{if}~ \forall x \in \mathit{xs}.\; \lnot\; P\; \mathit{xs} \\[-\jot]
&         \keyw{then}~\const{LNil} \\[-\jot]
&       \keyw{else}~\keyw{if}~ P\; (\hd\;\mathit{xs}) \\[-\jot]
&\quad         \keyw{then}~\const{LCons}\; (\hd\; \mathit{xs})\; (\const{lfilter}\; P\; (\tl\; \mathit{xs})) \\[-\jot]
&\quad       \keyw{else}~\const{lfilter}\; P\; (\tl\; \mathit{xs})\end{aligned}$
\end{quote}
The first self-call is corecursive and guarded by \const{LCons}, whereas the second
self-call is terminating, because the number of ``false'' elements
until reaching the next ``true'' element (whose existence is guaranteed by the
first \keyw{if} condition) decreases by one.

%
%
%
%
%
%
%
%

\subsection{Coinduction Up-to}
\label{sec-coind-exa}

Once a corecursive specification
has been accepted as productive, we normally want to reason about it. In proof
assistants, codatatypes are accompanied by a notion of structural
coinduction that matches primitively corecursive functions.
For nonprimitive specifications, our framework provides the more advanced
proof principle of
{coinduction up to congruence}---or simply \emph{coinduction up-to}.

The structural coinduction principle for streams is as follows:
\begin{center}
\AXC{$R\;l\;r$\,}
\AXC{$\forall s\;t.\;\, R\;s\;t \longrightarrow \hd\;s=\hd\;t \mathrel\land R\;(\tl\;s)\;(\tl\;t)$}
\BIC{$l = r$}
\DP
\end{center}
Coinduction allows us to prove an equality on streams by providing a relation
$R$ that relates the left-hand side with the right-hand side (first premise) and
that constitutes a bisimulation (second premise). Streams that are related by a
bisimulation cannot be distinguished by taking observations (i.e., by applying
the $\hd$ and $\tl$ selectors); therefore they must be equal. In other words,
equality is the largest bisimulation.

Creativity is generally required to instantiate $R$ with a bisimulation.
However, given a goal $l = r$, the following canonical candidate
often works: $\lambda s\;t.\;\exists\vthinspace\overline{\xs}.~s = l \mathrel\land
t = r$, where $\overline{\xs}$ are variables occurring free in $l$ or $r$.
%
As a rehearsal, let us 
prove that the primitively corecursive
operation $\opls$ is commutative.

\begin{prop}
$\xs \opls \ys = \ys \opls \xs$.
\end{prop}

\begin{proofx}
  We first show that $R = (\lambda s\;t.\;\exists\xs\;\ys.~ s = \xs \opls
  \ys \mathrel \land t = \ys \opls \xs)$ is a bisimulation. We fix two
  streams $s$ and $t$ for which we assume $R\;s\;t$ (i.e.,
  there exist two 
  streams $\xs$ and $\ys$ such that
  $s = \xs \opls \ys$ and $t = \ys \opls \xs$).
  Next, we must show that $\hd\;s = \hd\;t$ and $R\;(\tl\;s)\;(\tl\;t)$.
  The first property can be discharged by a simple calculation. For the second
  one:
  \begin{quote}
  $
  \begin{array}{@{}l@{}}
    R\;(\tl\;s)\;(\tl\;t)\\
    \iff R\;(\tl\;(\xs \opls \ys))\;(\tl\;(\ys \opls \xs))\\
    \iff R\;(\tl\;\xs \opls \tl\;\ys))\;(\tl\;\ys \opls \tl\;\xs)\\
    \iff\exists \xs'\;\ys'.\;\,\tl\;\xs \opls \tl\;\ys = \xs' \opls \ys' \mathrel\land {}\\
    \phantom{iff \exists \xs'\;\ys'.\;\,}\tl\;\ys \opls \tl\;\xs = \ys' \opls \xs'
  \end{array}
  $
  \end{quote}
  The last formula can be shown to hold by selecting $\xs' = \tl\;\xs$ and $\ys' =
  \tl\;\ys$. Moreover, $R\;(\xs \opls \ys)\;(\ys \opls \xs)$ holds. Therefore,
  the thesis follows by structural coinduction.
\end{proofx}

If we attempt to prove the commutativity of $\oprd$ analogously,
we eventually encounter
a formula of the form $R\;(\cdots\opls\cdots)\;(\cdots\opls\cdots)$, because
$\oprd$ is defined in terms of $\opls$. Since $R$ mentions only $\oprd$ but
not $\opls$, we are stuck. An ad hoc solution would be
to replace the canonical $R$ with a bisimulation that allows for descending
under $\opls$. However, this would be needed for almost every property
about~$\oprd$.

A more reusable solution is to strengthen the coinduction principle upon
registration of a new well-behaved operation. The strengthening mirrors the
acquired possibility of the new operation to appear in the corecursive call
context. It is technically represented by a congruence closure $\CONG : (\Stream
\ra \Stream \ra\nobreak \Bool) \ra\allowbreak \Stream \to \Stream \to \Bool$. The
coinduction up-to principle is almost identical to structural
coinduction, except that the corecursive application of $R$ is replaced by
$\CONG\;R$:
\begin{center}
\AXC{$R\;l\;r$}
\AXC{$\forall s\;t.\,R\;s\;t \longrightarrow \hd\;s=\hd\;t \mathrel\land \CONG\;R\;(\tl\;s)\;(\tl\;t)$}
\BIC{$l = r$}
\DP
\end{center}
The principle evolves with every newly registered well-behaved operation in the
sense that our framework refines the definition of the congruence closure
$\CONG$.
(Strictly speaking, a fresh symbol $\CONG'$ is introduced each time.)
For example, after registering $\SCons$ and $\opls$, $\CONG\;R$ is the
least reflexive, symmetric, transitive relation containing~$R$ and satisfying
the rules
\begin{center}
\begin{tabular}{@{}cc@{}}
\AXC{$x=y$}
\AXC{$\CONG\;R\;\xs\;\ys$}
\BIC{$\CONG\;R\;(\SCons\;x\;\xs)\;(\SCons\;y\;\ys)$}
\DP&

\AXC{$\CONG\;R\;\xs\;\ys$}
\AXC{$\CONG\;R\;\xs'\;\ys'$}
\BIC{$\CONG\;R\;(\xs \opls \xs')\;(\ys \opls \ys')$}
\DP
\end{tabular}
\end{center}
After defining and registering $\oprd$, the relation $\CONG\;R$ is
extended to also satisfy 
\begin{center}
\begin{tabular}{@{}c@{}}
\AXC{$\vphantom{x}\smash{\CONG\;R\;\xs\;\ys}$} 
\AXC{$\smash{\CONG\;R\;\xs'\;\ys'}$}
\BIC{$\CONG\;R\;(\xs \oprd \xs')\;(\ys \oprd \ys')$}
\DP
\end{tabular}
\end{center}

Let us apply the strengthened coinduction principle to prove the
distributivity of stream exponentiation over pointwise addition:

\begin{prop} 
$\oexp\;(\xs \opls \ys) = \oexp\;\xs \oprd \oexp\;\ys$.
\end{prop}
\begin{proofx}
  We first show that $R = (\lambda s\;t.\;\exists\xs\;\ys.~ s = \oexp\;(\xs \opls \ys) \mathrel\land t = \oexp\;\xs \oprd \oexp\;\ys)$ is a bisimulation.
  We fix two
  streams $s$ and $t$ for which we assume $R\;s\;t$ (i.e., there exist two 
  streams $\xs$ and $\ys$ such that $s = \oexp\;(\xs \opls \ys)$ and $t = \oexp\;\xs \oprd \oexp\;\ys$).
  Next, we show that $\hd\;s = \hd\;t$ and $\CONG\;R\;(\tl\;s)\;(\tl\;t)$:
  \begin{quote}
  $
  \begin{array}{@{}r@{\;}c@{\;}l@{}}
    \hd\;s &=& \hd\;(\oexp\;(\xs \opls \ys))
    = 2 \mathbin{\vthinspace\char`\^\vthinspace} \hd\;(\xs \opls \ys)\\
    &=& 2 \mathbin{\vthinspace\char`\^\vthinspace} (\hd\;\xs + \hd\;\ys)
    =  2 \mathbin{\vthinspace\char`\^\vthinspace} \hd\;\xs \times 2 \mathbin{\vthinspace\char`\^\vthinspace} \hd\;\ys\\
    &=& \hd\;(\oexp\;\xs) \times \hd\;(\oexp\;\ys)\\
    &=& \hd\;(\oexp\;\xs \oprd \oexp\;\ys) = \hd\;t
  \end{array}
  $\\[2\jot]
  $
  \begin{array}{@{}l@{}}
    \CONG\;R\;(\tl\;s)\;(\tl\;t)\\
    \iff \CONG\;R\;(\tl\;(\oexp\;(\xs \opls \ys)))\;(\tl\;(\oexp\;\xs \oprd \oexp\;\ys))\\
    \iff \CONG\;R\;((\tl\;\xs \opls \tl\;\ys) \oprd \oexp\;(\xs \opls \ys))\;\\
    \phantom{\iff}(\oexp\;\xs \oprd (\tl\;\ys \oprd \oexp\;\ys) \mathrel{\opls} (\tl\;\xs \oprd \oexp\;\xs) \oprd \oexp\;\ys)\\
    \smash{\stackrel{*}{\iff}}~\CONG\;R\;((\tl\;\xs \oprd \oexp\;(\xs \opls \ys) \opls \tl\;\ys \oprd \oexp\;(\xs \opls \ys))\;\\
    \phantom{\iff}(\tl\;\xs \oprd (\oexp\;\xs \oprd \oexp\;\ys) \mathrel{\opls} \tl\;\ys \oprd (\oexp\;\xs \oprd \oexp\;\ys)\\
    \smash{\stackrel{\opls}{\iffla}}~\CONG\;R\;(\tl\;\xs \oprd \oexp\;(\xs \opls \ys))\;(\tl\;\xs \oprd (\oexp\;\xs \oprd \oexp\;\ys)) \mathrel\land {}\\
    \phantom{\smash{\stackrel{\opls}{\iffla}}}~\CONG\;R\;(\tl\;\ys \oprd \oexp\;(\xs \opls \ys))\;(\tl\;\ys \oprd (\oexp\;\xs \oprd \oexp\;\ys))\\
    \smash{\stackrel{\oprd}{\iffla}}~\CONG\;R\;(\tl\;\xs)\;(\tl\;\xs) \mathrel\land \CONG\;R\;(\tl\;\ys)\;(\tl\;\ys) \mathrel\land \\
    \phantom{\smash{\stackrel{\oprd}{\iffla}}}~\CONG\;R\;(\oexp\;(\xs \opls \ys))\;(\oexp\;\xs \oprd \oexp\;\ys)\\
    \iffla R\;(\oexp\;(\xs \opls \ys))\;(\oexp\;\xs \oprd \oexp\;\ys)
  \end{array}
  $\kern-200mm 
  \end{quote}
  The step marked with $*$ appeals to associativity and commutativity of $\opls$ and
  $\oprd$ as well as distributivity of $\oprd$ over $\opls$. These properties are
  likewise proved by coinduction up-to. The implications marked with $\opls$
  and $\oprd$ are justified by the respective congruence rules. The last
  implication uses reflexivity and expands $R$ to its closure $\CONG\;R$.

  Finally, it is easy to see that $R\;(\oexp\;(\xs \opls \ys))\;(\oexp\;\xs \oprd
  \oexp\;\ys)$ holds. Therefore, the thesis follows by coinduction up-to.
\end{proofx}

The formalization accompanying this paper \cite{our-formalization} also
contains proofs of
$\facA = \facC\;1\;1\;1 = \mapS\;\fac\;(\const{natsFrom}\;1)$, $\facB = \SCons\;1\;\facA$, and
$\fibA = \fibB$,
where $\fac$ is the factorial on $\Nat$.

Nested corecursion up-to is also reflected with a suitable
strengthened coinduction rule.
For $\Tree$, this strengthening takes place
under the $\listrel$ operator on list, similarly to the corecursive calls
occurring nested in the $\map$ function:
\begin{center}
\AXC{\kern-1pt\strut$R\;l\;r$} 
\AXC{\strut$\forall s\;t.\,R\;s\;t \longrightarrow \Label\;s=\Label\;t \mathrel\land \listrel\;(\CONG\;R)\;(\Children\;s)\;(\Children\;t)$\kern-1pt} 
\BIC{$l = r$}
\DP
\end{center}
The $\listrel\;R$ operator lifts the binary predicate $R : A \ra B \ra \Bool$ to
a predicate $\List\;A \ra \List\;B \ra \Bool$. More precisely,
$\listrel\;R\;\xs\;\ys$ holds if and only if $\xs$ and $\ys$ have the same length
and parallel elements of $\xs$ and $\ys$ are related by
$R$. This nested coinduction rule is convenient provided there is
some infrastructure to descend under $\listrel$ (as is the case
in Isabelle\slash HOL). The formalization~\cite{our-formalization} establishes
several arithmetic properties of $\boxplus$ and~$\boxtimes$.

\section{Extensible Corecursors}
\label{sec-meta}

We now describe the definitional and proof mechanisms that substantiate flexible corecursive definitions
in the style of \S\ref{sec-exa}.  They are based on the modular maintenance of
infrastructure for the corecursor associated with a codatatype, with the possibility of open-ended incremental improvement.
We present the approach for an arbitrary codatatype
given as the greatest fixpoint of an arbitrary (bounded) functor.
The approach is quite general
and does not rely on any particular grammar for specifying codatatypes.

Extensibility is an integral feature of the framework. In principle, an
implementation could redo the constructions from scratch each time a
well-behaved operation is registered, but it would give rise to a quadratic
number of definitions, slowing down the proof assistant. The incremental
approach is also more flexible and future-proof, allowing mixed fixpoints and
composition with other (co)recursors, including some that do not exist yet.


\subsection{Functors and Relators}
\label{sec-funcs-rels}

Functional programming languages and proof assistants
necessarily maintain a database of the user-defined
types or, more generally, type constructors, which
can be thought as functions $\F : \Set^n \ra \Set$ operating on sets (or perhaps on ordered sets).
It is often useful to maintain more structure along with these type constructors:
\begin{itemize}
\item a functorial action $\Fmap : \smash{\prod_{\ov{A},\ov{B} \in \Set^n} \prod_{i = 1}^n (A_i \ra\nobreak B_i)} \ra\allowbreak \F\;\ov{A} \ra \F\;\ov{B}$, i.e., a polymorphic function
of the indicated type that commutes with identity 
$\id_A : A \ra A$ and composition;
\item a relator $\Frel : \smash{\prod_{\ov{A},\ov{B} \in \Set^n} \prod_{i = 1}^n (A_i \ra B_i \ra \Bool)} \ra \F\;\ov{A} \ra \F\;\ov{B} \ra \Bool$, i.e.,
a polymorphic function of the indicated type which commutes with binary-relation identity and composition.
\end{itemize}
Following standard notation from category theory, we write $\F$ instead of $\Fmap$.
%
Given binary relations $R_i : A_i \ra B_i \ra \Bool$ for $1\leq i \leq n$, we
think of $\Frel\;\ov{R} : \F\;\ov{A} \ra \F\;\ov{B} \ra \Bool$ as the natural
lifting of $R$ along $\F$; for example, if $\F$ is $\List$ (and hence $n = 1$),
$\Frel$ lifts a relation on elements to the componentwise relation on lists (also requiring equal length). 
It is well known that the positive type constructors defined by standard means (basic types, composition, least or greatest fixpoints)
have canonical functorial and relator structure. This is crucial
for the foundational construction of user-specified (co)datatypes in Isabelle\slash HOL  \cite{traytel-et-al-2012}.

But even nonpositive type constructors $\G : \Set^n \ra \Set$
exhibit a relator-like structure $\Grel : \smash{\prod_{\ov{A},\ov{B} \in \Set^n}} \;(\ov{A} \ra \ov{B}) \ra (\G\,\ov{A} \ra \G\,\ov{B} \ra \Bool)$ (which
need not commute with relation composition, though).
For example, if $\G : \smash{\Set^2} \ra \Set$ is the function-space constructor $\G\;(A_1,A_2) = A_1 \ra A_2$ and
$f \in \G\;(A_1,A_2)$, $g \in \G\;(B_1,B_2)$,
$R_1 : A_1 \ra B_1 \ra \Bool$, and $R_2 : A_2 \ra B_2 \ra \Bool$,
then $\Grel\;R_1\;R_2\;f\;g$ is defined as $\forall a_1 \in A_1.\; \forall b_1 \in B_1.\;R_1\;a_1\;b_1 \lra R_2\;(f\;a_1)\;(g\;b_1)$.
A polymorphic function $c : \prod_{\ov{A} \in \Set^n} \G\;\ov{A}$, $c$ is called {\em parametric} \cite{rey-param,wadler-89}
if $\forall \ov{A},\ov{B} \in \Set^n.\;\forall R : \ov{A} \ra \ov{B} \ra \Bool.\;\Grel\;R\;c_{\ov{A}}\;c_{\ov{B}}$.
%
The maintenance of relator-like structures is very helpful for automating theorem transfer along isomorphisms and quotients
\cite{huffman-lifting}.
Here we explore an additional benefit of maintaining functorial and relator structure for type constructors:\
the possibility to 
extend the corecursor in reaction to user input.

In this section, we assume that all the considered type constructors are both functors and relators,
that they include
basic functors such as identity, constant, sum, and product, and that they are closed under
least fixpoints (initial algebras) and greatest fixpoints (final coalgebras).
Examples of such classes of type constructors include the datafunctors \cite{hensel-interatedRecursion}, the containers \cite{abbott-et-al-2005},
and the bounded natural functors \cite{traytel-et-al-2012}. 

\leftOut{
\begin{figure}
\small
$$
\xymatrix@C=.1333pc@R=.2pc{
 & & & & & & *=0{} \ar@/^0pt/@{}[ddddddllllll] \ar@/^0pt/@{}[ddddddrrrrrr] \\
\\
\\
&&&&&& \Smash{\kern.8em\XDot a_2} \\
&&&& \Smash{\kern.8em\XDot a_1} \\
&&&&&&&& \Smash{\kern.8em\XDot a_3} \\
 & & & & & & & & & & & &
}
$$
\vspace*{-2ex} 
\caption{An element $x$ of $\F\;A$ with content items $a_1,a_2,a_3$}
\label{fig-elem}
\end{figure}
}

We focus on the case of a unary codatatype-generating functor $\F : \Set \ra \Set$.
The codatatype of interest will be its greatest fixpoint
(or final coalgebra)
$\J = \gfp\;\F$.
This generic situation
already covers the vast majority of interesting codatatypes, 
since $\F$ can represent arbitrarily complex nesting.
For example,
if $\F = (\lambda A.~\Nat \times \List\;A)$, then $\J$ corresponds to the $\Tree$ codatatype
presented in Section~\ref{sec-nest-exa}.
The extension to mutually defined codatatypes is straightforward but tedious.
Our examples will take $\J$ to be the $\Stream$ type from Section~\ref{sec-exa}, with
$\F = (\lambda A.~ \Nat \times A)$.

Given a set $A$,
it will be useful to think of the elements $x \in \F\;A$ as consisting of a {\em shape} together with
{\em content} that fills the shape with elements of $A$.
%
If $\F\;A = \Nat \times A$, the shape of $x = (n,a)$ is $(n,\_)$ and the content is $a$;
if $\F\;A = \List\;A$, the shape of $x = [x_1,\ldots ,x_n]$ is the $n$-slot
container $[\_,\ldots,\_]$ and the content consists of the $x_i$'s.

According to this view,
for each $f : A \ra B$, the functorial action associated with $\F$ 
sends any $x$
into an element $\F\;f\;x$ of the same shape as~$x$ but with each content item $a$ replaced by $f\;a$.
Technically, this view
can be supported by custom notions such as
containers \cite{abbott-et-al-2005} or, 
more simply, via a parametric 
function of type $\prod_{A \in \Set} \F\;A \ra \Set\;A$ that collects the content elements \cite{traytel-et-al-2012}.

\subsection{Primitive Corecursion}
\label{sec-prim}

%
%
The \keyw{codatatype} that defines $\J$
also introduces the constructor and destructor bijections
$\ctor : \F\,\J \ra \J$ and $\dtor : \J \ra \F\,\J$ and the primitive corecursor
$\corec : \prod_{A \in \Set} (A \ra \F\;A) \ra A \ra \J$ characterized
by the equation 
$\corec\;s\;a = \ctor\,(\F\,(\corec\;s)\;(s\,a))$.
In elements $x \in \F\;A$,
the occurrences of content items $a \in A$ in the shape of $x$
captures the positioning of the corecursive calls.

\begin{eexample}\label{exa-prim} \rm
Modulo currying, the pointwise sum of streams $\opls$ is definable as $\corec\;s$,
by taking $s : \Stream^2 \ra \Nat \times \Stream^2$ to be $\lambda (\xs,\ys).~(\hd\;\xs + \hd\;\ys, (\tl\;\xs,\tl\;\ys))$.
\end{eexample}

In Example~\ref{exa-prim} and elsewhere, we lighten notation by identify
the curried and uncurried forms of functions, counting on implicit coercions
between the two.

\subsection{The Corecursion State}
\label{sec-state}

Given any functor $\SS : \Set \ra \Set$, we define its {\em free-monad functor} $\SS^*$ by $\SS^*A = \lfp\,(\lambda B.~A + \SS\;B)$.
We write $\eta : A \ra \SS^*A$ and $\talg : \SS\,(\SS^*A) \ra \SS^*A$ for the left and right injections
into $\SS^*A$.

The functions $\eta$ and $\talg$ are in fact polymorphic; for example,
$\eta$ has type $\prod_{A \in \Set} A \ra \SS^*A$.
We often omit the set parameters of polymorphic functions if they can be inferred from the context, writing $\eta$ and $\talg$ instead of
$\eta_A$ and $\talg_A$.


At any given moment, we maintain the following data associated with $\J$, which we call a
{\em corecursion state}: 
\begin{itemize}
\item a finite number of functors $\K_1,\ldots,\K_n : \Set \ra \Set$ and, for each $\K_i$,
a function $f_i : \K_i\;\J \ra \J$;
\item a polymorphic function $\l : \prod_{A \in \Set} \SS\,(A \times \F\;A) \ra \F\,(\SS^*A)$.
\end{itemize}
We call the $f_i$'s the {\em well-behaved operations} and define their
collective {\em signature functor} $\SS$
as $\lambda A.~\K_1\;A + \cdots + \K_n\;A$, where $\iota_i : \K_i \ra \SS$
is the standard embedding of $\K_i$ into $\SS$.
We call $\l$ the {\em corecursor seed}.
The corecursion state is subject to the following conditions:
\begin{description}
\item[Parametricity:] $\l$ is parametric.
\item[Well-behavedness:] Each $f_i$ satisfies the characteristic equation
$$f_i\;x =  \ctor\,(\F\;\eval\;(\l\,(\SS\;\langle\id,\dtor\rangle\,(\iota_i\,x)))) $$
\end{description}
The convolution operator $\langle \_,\_ \rangle$ builds a function
$\langle f,g \rangle : B  \ra C \times D$ from
two functions $f : B \ra C$ and $g : B \ra D$,
and $\eval : \SS^*\J \ra \J$ is the canonical
evaluation function defined recursively (using the primitive recursor associated with $\SS^*$):
\begin{quote}
$\eval\;(\eta\;j)  =  j$\\
$\rlap{$\eval\,(\talg\;z)$}\phantom{\eval\;(\eta\;j)}  =  \keyw{case}~z~\keyw{of}~\iota_i\;t \Ra f_i\;(\K_i\,\eval\;t)$\kern-\leftmargin
\end{quote}
(Note that, on the recursive call on the right $\eval$, is applied to $t$ via ``lifting'' it through the functor $K_i$.)
Functions having the type of $\l$
and additionally assumed parametric (or, equivalently, assumed to be natural transformations)
are known in category theory as ``abstract GSOS rules.''
They were introduced by Turi and Plotkin \cite{turi-plotkin97}
and further studied by Bartels \cite{BartelsGeneralizedCoind}, Jacobs \cite{jacobs06-distrib},
Hinze and James \cite{hinze-adventure},
Milius et al.\ \cite{milius-modular}, and others.


Thus, a corecursion state is a triple $(\ov{\K},\ov{f},\l)$.
As we will see in \S\ref{sec-adv}, the state {\em evolves} as users
define and register new functions.
%
The $f_i$'s are the operations that have been registered as
safe for participating in the context of corecursion calls.
Since $f_i$ has type $\K_i\;\J \ra \J$, we think of $\K_i$ as encoding the arity of $f_i$.
Then $\SS$, the sum of the $\K_i$'s, represents the signature consisting of all the $f_i$'s.
Thus, for each $A$,
$\SS^*A$ represents the set of formal expressions over $\Sigma$ and $A$, i.e., the trees built
starting from the ``variables'' in $A$ as leaves by
applying operations symbols corresponding to the $f_i$'s.
Finally, $\eval$ evaluates in $\J$ the formal expressions of $\SS^*\J$ by applying
the functions $f_i$ recursively.

If the functors $\K_i$ are restricted to be finite monomials $\lambda A.\;A^{k_i}$, the functor
$\SS$ can be seen as a standard algebraic signature
and $(\SS^*A,\vthinspace \talg)$ as the standard term algebra for this signature, over the variables $A$.
However, we allow $\K_i$ to be more exotic; for example, $\K_i\,A$
can be $A^\Nat$ (representing an infinitary operation) or one of $\List\;A$ and $\FSet\;A$
(representing an operation taking a varying finite number of ordered or unordered arguments).

But what guarantees that the $f_i$'s are indeed safe as contexts for corecursive
calls? In particular, how can the framework
exclude $\tl$ while allowing $\SCons$, $\opls$, and
$\otimes$? This is where the parametricity and well-behavedness conditions on
the state enter the picture.

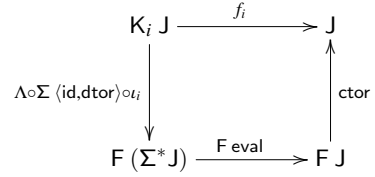
\begin{figure}
$$
\xymatrix@C=3.5pc@R=3pc{
 \K_i\;\J
\ar^{f_i}[r]
\ar_{\l \mathrel\circ \SS\,\langle\id,\dtor\rangle \mathrel\circ \iota_i}[d]    &  \J
                   \\
\F\,(\SS^*\J)
\ar^{\F\;\eval}[r]  &  \F\;\J \ar_{\ctor}[u]
}
$$
\vspace*{-1ex} 
\caption{The well-behavedness condition}
\label{fig-reg}
\end{figure}

We start with well-behavedness.
Assume $x \in \K_i$, which is unambiguously represented in $\SS$ as $\iota_i\;x$.
Let $j_1, \ldots, j_m \in \J$ be the content items of $\iota_i\;x$ (placed in various slots
in the shape of $x$). 
To evaluate $f_i$ on $x$,
we first corecursively destruct the $j_l$'s while also keeping the originals, thus replacing each $j_l$ with
$(j_l,\vthinspace \dtor\,j_l)$.
Then we apply the transformation $\l$ to obtain an element of $\F\,(\SS^*\J)$, which has an $\F$-shape at the top
(the first produced observable data) and for each slot in this shape an element of $\SS^*\J$,
i.e., a formal-expression tree having leaves in $\J$ and built using operation symbols from the signature
(the corecursive continuation):
%
$$\K_i\;\J \stackrel{\iota_i}{\lra} \SS\;\J \stackrel{\SS\,\langle\id,\dtor\rangle}{\lra} \SS\;(\J \times \F\;\J) \stackrel{\l}{\lra} \F\,(\SS^*\J)$$
In summary, $\l$ is a schematic representation of the mutually corecursive behavior of the well-behaved operations
up to the production of the first observable data.
This intuition is made formal in the well-behavedness condition, which states that the diagram in Figure~\ref{fig-reg} commutes for each $f_i$. (We could replace the right upward arrow labeled by $\ctor$ with a downward arrow labeled by $\dtor$ without changing the diagram's meaning.
However, we consistently prefer the constructor view in our exposition.)

In the above explanations, we saw that it suffices to peel off one layer of
the arguments $j_i$ (by applying 
$\dtor$)
for a well-behaved operation
$f_i$ to produce, via $\l$, one layer of the result and to delegate the rest of the computation to a context consisting of a combination of well-behaved
operations (an element of $\SS^*\J$). But how to formally express that exploring one layer is enough, i.e., that applying $\l : \J \times \F\;\J \ra \F\,(\SS^*\J)$
to $(j_i,\vthinspace\dtor\,j_i)$
does not result in a deeper exploration? An elegant way of capturing this is to require that $\l$, which is a polymorphic function,
operates without analyzing $\J$, i.e., that it
operates in the same way on $A \times \F\;A \ra \F\,(\SS^*A)$ for any set $A$. This
requirement is precisely parametricity.

Strictly speaking, the well-behaved operations $\ov{f}$ are a redundant piece of data in the state $(\ov{\K},\ov{f},\l)$,
since, assuming $\l$ parametric, we can prove that there exists a unique tuple $\ov{f}$ that satisfies the well-behavedness condition.
In other words, the operations $\ov{f}$ could be derived on a per-need basis.


\begin{eexample}\rm \label{exa-state}
Let $\J = \Stream$ and assume that $\SCons : \Nat \times \Stream \ra \Stream$ and
$\opls : \Stream^2 \ra \Stream$ are the only well-behaved operations registered so far.
Then $\K_1 = (\lambda B.~\Nat \times\nobreak B$), $f_1 = \SCons$,
$\K_2 = (\lambda B.\;B^2)$, and $f_2 = {\opls}$.
Moreover, $\SS^* = \lfp\;(\lambda B.\allowbreak~A \allowbreak + (\Nat \times B + B^2))$
consists of formal-expression trees with leaves in $A$ and built using
arity-correct applications of operation symbols corresponding to
$\SCons$ and $\opls$, denoted by $\bb{\SCons}$ and $\bb{\opls}$.  Given $n \in \Nat$ and $a,b \in A$,
an example of such a tree is $\eta\;a \;\bb{\opls}\; \bb{\SCons}\,(n,\eta\;a \,\bb{\opls}\, \eta\;b)$.
If additionally $A = \J$, then $\eval$ applied to the above tree is $a \,\opls\, \SCons\;n\;(a \,\opls\, b)$.

But what is $\l$? As we show below, we need not worry about the global definition of $\l$, since both $\SS$ and $\l$ will be updated
{incrementally} when registering new operations as well behaved.
Nonetheless, a global definition of $\l$ for $\SCons$ and $\opls$ follows:
$$
\begin{array}{@{}l@{}}
\l\;z = \textsf{case $z$ of} \\
\hspace*{1.3ex}
 \begin{array}{@{}l@{\;}l@{\;}l@{}}
  \iota_1\,(n,(a,(m,a')) & \Ra & (n,\vthinspace \bb{\SCons}\vthinspace(m,\eta\;a')) \\
  \iota_2\,((a,(m,a')),\;(b,(n,b'))) & \Ra & (m + n,\vthinspace \eta\;a' \,\bb{\opls}\, \eta\;b')
  \end{array}
\end{array}
$$
%
Informally, $\SCons$ and $\opls$ exhibit the following behaviors:
\begin{itemize}
\item to evaluate $\SCons$ on a number $n$ and an item $a$ with $(\hd\;a,\allowbreak \tl\;a) = (m,a')$,
produce $n$ and evaluate $\SCons$ on $m$ and $a'$, i.e.,
output $\SCons\;n\;(\SCons\;m\;a') = \SCons\;n\;a$;
\item to evaluate $\opls$ on 
$a, b$ with $(\hd\;a, \tl\;a) = (m,a')$ and $(\hd\;b,\tl\;b) = (n,b')$,
produce $m+n$ and evaluate $\opls$ on $a'$ and $b'$, i.e., output
$\SCons\;(m+n)\;(a' \opls b')$.
\end{itemize}
%
\end{eexample}

\subsection{Corecursion Up-to}
\label{sec-corec-princ}

A corecursion state $(\ov{\K},\ov{f},\l)$ for an $\F$-defined codatatype $\J$ consists of a collection of operations
on $\J$, $f_i : \K_i\;\J \ra \J$, that satisfy the well-behavedness properties
expressed in terms of a parametric function $\l$.
We are now ready to harvest the crop of this setting:\ a corecursion principle
for defining functions having $\J$ as codomain.

\begin{figure*}
        \centering
        \begin{subfigure}[b]{0.175\textwidth}
\centering
$\xymatrix@C=3.5pc@R=3pc{
 A
\ar^{\corec\;s}[r]
\ar_{s}[d]    &  \J
                   \\
\F\;A
\ar^{\F\,(\corec\;s)}[r]  &  \F\;\J \ar^{\ctor}[u]
}$
                \caption{Primitive corecursion}
                \label{fig-prim-corec}
        \end{subfigure}%
        \quad
        \begin{subfigure}[b]{0.325\textwidth}
\centering
$\xymatrix@C=2pc@R=2pc{
 A
\ar^{\corecU\;s}[rr]
\ar_{s}[d]    & &  \J
                   \\
\F\;(\SS^*A)
\ar_{\!\!\!\!\F\,(\SS^*\,(\corecU\;s))\;\;}[rd]  &   & \F\;\J \ar^{\ctor}[u] \\
  & \F\;(\SS^*\J) \ar^{\F\;\eval}[ru] &
}$
                \caption{Top-guarded corecursion up-to}
                \label{fig-top-guarded-corec}
        \end{subfigure}
        \quad
        \begin{subfigure}[b]{0.425\textwidth}
\centering
$\xymatrix@C=2.5pc@R=2pc{
 A
\ar^{\corecUU\;s}[rr]
\ar_{s}[d]    & &  \J
                   \\
\SS^*\,(\F\;(\SS^*A))
\ar_{\SS^* (\F\,(\SS^*\,(\corecUU\;s)))\;\;\;\;\;\;\;\;\;\;}[dr]  &   & \SS^*\J \ar^{\eval}[u] \\
   & \SS^*\,(\F\,(\SS^*\J)) \ar^{\hspace*{2ex}\SS^*\,(\F\;\eval)}[r] & \SS^*\,(\F\;\J) \ar^{\SS^*\;\ctor}[u]
}$
                \caption{Flexibly guarded corecursion up-to}
                \label{fig-flexibly-guarded-corec}
        \end{subfigure}
        \caption{The corecursors}\label{fig-corec}
\end{figure*}
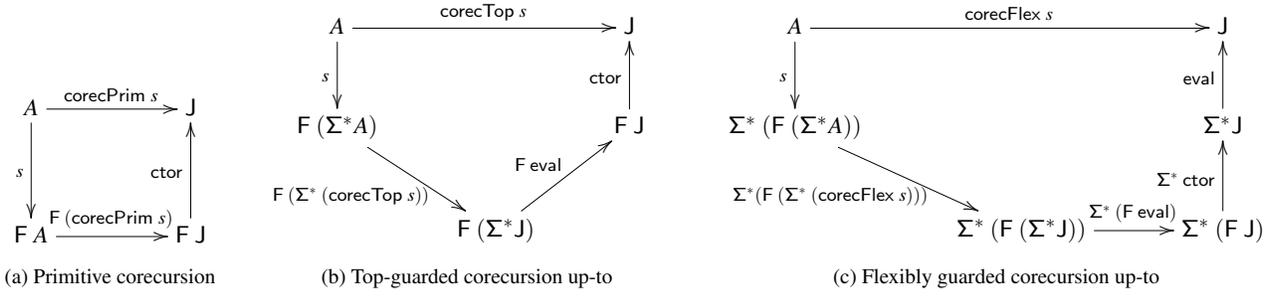

The principle will be represented by two corecursors, $\corecU$ and $\corecUU$. Although
subsumed by the latter, the former is interesting in its own right and will
give us the opportunity to illustrate some fine points. Below
we list the types of these corecursors
along with that of the primitive corecursor for comparison:
\begin{description}
\item[Primitive corecursor:]
\ \\\hspace*{3ex}
$\corec : \prod_{A \in \Set} (A \ra \F\;A) \ra A \ra \J$
\item[Top-guarded corecursor up-to:]
\ \\\hspace*{3ex}
$\corecU : \prod_{A \in \Set} (A \ra \F\,(\SS^*A)) \ra A \ra \J$
\item[Flexibly guarded corecursor up-to:]
\ \\\hspace*{3ex}
$\corecUU : \prod_{A \in \Set} (A \ra \SS^*\,(\F\,(\SS^*A))) \ra A \ra \J$
\end{description}
Figure~\ref{fig-corec} presents the diagrams whose commutativity properties give
the characteristic equations of these corecursors.

Each corecursor implements a contract of the following form:
If, for each $a \in A$, one provides the intended corecursive behavior of $g\;a$ represented as $s\;a$,
where $s$ is a function
from~$A$,
one obtains the function $g : A \ra \J$ (as the corresponding corecursor applied to $s$) satisfying a suitable fixpoint equation matching this behavior.

The codomain of $s$ is the key to understanding the expressiveness of each corecursor.
The intended corecursive calls are represented by $A$,
and the call context is represented by the surrounding
combination of functors (involving $\F$, $\SS^*$, or both):
\begin{itemize}
\item for $\corec$, the allowed call contexts consist of a single constructor guard (represented by~$\F$);
\item for $\corecU$, they consist of a constructor guard (represented by~$\F$) followed by any combination of well-behaved
operations $f_i$ (represented by $\SS^*$);
\item for $\corecUU$, they consist of any combination of well-behaved operations
satisfying the condition that on every path leading to a corecursive call there exists at least one constructor guard
(represented by $\SS^*\,(\F\,(\SS^*\_))$).
\end{itemize}
We can see the computation of $g\;a$
by following the diagrams in Figure~\ref{fig-corec} counterclockwise from their left-top corners.
The application $s\;a$ first builds the call context {syntactically}.
Then $g$ is applied corecursively on the leaves.
Finally, the call context is evaluated:\ for $\corec$, it consist only of the guard ($\ctor$);
for $\corecU$, it involves the evaluation of the well-behaved operators (which may also include several occurrences of the guard)
and ends with
the evaluation of the top guard;
for $\corecUU$, the evaluation of the guard is
interspersed with that of the other well-behaved operations.

\begin{eexample}\rm \label{exa-defUpToInst}
For each example from \S\ref{sec-uptoExa}, we give
the corecursors that can handle it (assuming the
necessary well-behaved operations were registered):
$$
\begin{array}{l@{\enskip}l}
\opls, \everyOther{:} & \corecUU, \corecU, \corec \\
\onetwos, \fibA, \oprd, \exp, {\sup}{:} & \corecUU, \corecU   \\ 
\fibB, \facA, \facB{:} & \corecUU
\end{array}
$$
With the usual identification of $\Unit \ra \J$ and $\J$, we can define
$\fibA$ and $\facA$ as follows:
\begin{quote}
$\fibA  =  \corecU\,(\lambda u:\Unit.~(0,\vthinspace\bb{\SCons}(1,\eta\;u) \,\bb{\opls}\, (\eta\;u)))$\kern-200mm 
\\[1\jot]
$\facA  =  \corecUU\,(\lambda u:\Unit.~\eta\;(1,\eta\;u) \,\bb{\oprd}\, \eta\;(1,\eta\;u))$\kern-200mm 
\end{quote}
%
%
%
Let us look at $\fibA$ 
closely, comparing its specification
$\fibA  =  \SCons\;0\;(\SCons\;1\;\fibA \vvthinspace$ $\opls\vvthinspace \fibA)$
with its definition in terms of $\corecU$.
The outer $\SCons$ guard (with $0$ as first argument) corresponds to the outer
pair $(0,\_)$. The inner $\SCons$ and $\opls$ are interpreted as well-behaved operations
and represented by the symbols $\bb{\SCons}$ and $\bb{\opls}$ (cf.\ Example \ref{exa-state}).
Finally, the corecursive calls of $\fibA$ are captured by $\eta\;u$.

The desired specification can be obtained from the $\corecU$ form by the characteristic equation
of $\corecU$ (for $A = \Unit$) and the properties of $\eval$ as follows, where we simply write $s$, $\fibA$, and $\Call$ instead of their applications
to the unique element $()$ of $\Unit$,
namely $s\,()$, $\fibA\,()$, and $\Call\,()$:
$$
  \begin{array}{@{}l@{}}
    \phantom{{=}\;}\fibA \\
    {=}\hfill\mbox{\{by the commutativity of Figure~\ref{fig-top-guarded-corec}\phantom{\}}}\\
	\phantom{=}\hfill\mbox{\phantom{\{}with \ensuremath{\fibA = \corecU\;s}\}} \\
    \phantom{{=}\;}\ctor\,(\F\,(\eval \mathrel\circ \SS^*\;\fibA)\,s) \\
    {=}\hfill\mbox{\{by the definitions of $\F$ and $s$\}}\\
    \phantom{{=}\;}\SCons\;0\;((\eval \mathrel\circ \SS^*\;\fibA)\;
      (\bb{\SCons}(1,\Call) \,\bb{\opls}\, (\Call))) \\
    {=}\hfill\mbox{\{by the definition of $\Sigma^*$\}}\\
    \phantom{{=}\;}\SCons\;0\;(\eval \, (\bb{\SCons}(1,\Call\;\fibA) \,\bb{\opls}\, (\Call\;\fibA))  \\
    {=}\hfill\mbox{\{by the definition of $\eval$\}} \\
    \phantom{{=}\;}\SCons\;0\;(\SCons\;1\;\fibA \vvthinspace\opls\vvthinspace \fibA)
  \end{array}
  $$
The elimination of the $\corecU$ infrastructure relies on
simplification rules for the involved operators
and can be fully automatized. 
%
%
\end{eexample}

Parametricity and well-behavedness are crucial for proving that the corecursors
actually exist:

\begin{theorem}\label{corec-char}\rm
There exist the polymorphic functions $\corecU$ and $\corecUU$
making the diagrams in Figures~\ref{fig-top-guarded-corec} and \ref{fig-flexibly-guarded-corec} commute.
Moreover, for each $s$ of appropriate type, $\corecU\;s$
or $\corecUU\;s$ is the unique
function making its diagram commute.
\end{theorem}

Theorem \ref{corec-char} is a known result from the category theory literature:
The $\corecU\;s$ version follows from the results in Bartels's thesis \cite{bartels-thesis}, whereas the $\corecUU\;s$ version
was very recently (and independently) proved by Milius et al.\ \cite[Theorem~2.16]{milius-modular}.

\subsection{Initializing the Corecursion State}
\label{sec-init}

The simplest relaxation of primitive corecursion
is the allowance of multiple constructors in the call context,
in the style of Coq,
as in the definition of $\onetwos$ (\S\ref{sec-uptoExa}).
Since this idea is independent of the choice of
codatatype $\J$, we realize it when bootstrapping the corecursion state.

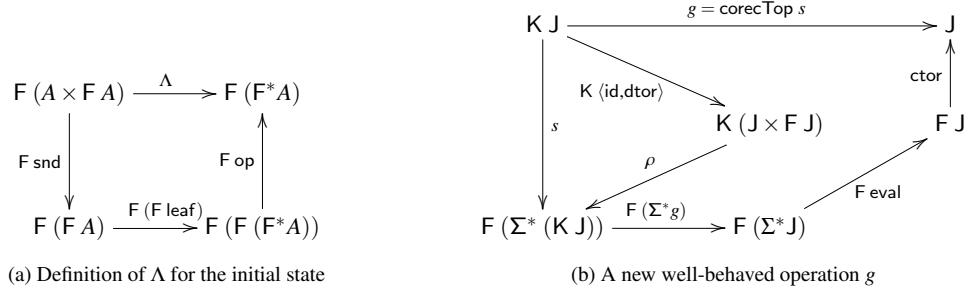
\begin{figure*}[t]\centering
\hspace*{-5em}
\begin{subfigure}[b]{0.4\textwidth}
$$
\xymatrix@C=2pc@R=3pc{
 \F\,(A \times \F\;A)
\ar^{\l}[r]
\ar_{\F\;\snd}[d]    &  \F\,(\F^*A)
                   \\
\F\,(\F\;A)
\ar^{\F\;(\F\;\eta)}[r]  &  \F\;(\F\;(\F^*A)) \ar^{\F\;\talg}[u]
}
%
%
%
%
$$
\vspace*{-1ex} 
\caption{Definition of $\l$ for the initial state} 
\label{fig-init}
\end{subfigure}\quad
\begin{subfigure}[b]{0.4\textwidth}
\centering
$\xymatrix@C=3pc@R=2pc{
 \K\;\J \ar^{s}[dd] \ar^{g \;=\; \corecU\;s}[rr]\ar_{\K\,\langle\id, \dtor\rangle\,\,\,\,}[dr]    & &  \J
\\
 & \K\,(\J \times \F\;\J) \ar_{\rho}[dl]           & \F\;\J \ar^{\ctor}[u]
\\
  \F\;({\SS}^*\,(\K\;\J))\ar^{\F\,(\SS^* g)}[r]     & \F\;({\Sigma}^*\J) \ar_{\F\;\eval}[ur]  &
}$
\caption{A new well-behaved operation $g$}
\label{fig-integg}
\end{subfigure}
\caption{Definitions of $\l$ and $g$}
\label{fig-init-integg}
\end{figure*}

More precisely, upon defining a codatatype $\J$, we take the following initial corecursion state $\initState = (\ov{\K},\ov{f},\l)$:
\begin{itemize}
\item $\ov{K}$ is a singleton consisting of (a copy of) $\F$; 
\item $\ov{f}$ is a singleton consisting of $\ctor$;
\item $\l : \prod_{A \in \Set} \F\,(A \times \F\;A) \ra \F\,(\F^*\;A)$ is defined as $\F\,(\talg \mathrel\circ \F\;\eta \mathrel\circ \snd)$,
where $\snd$ is the second product projection. 
\end{itemize}
Recall that the seed $\l$ is designed to
schematically represent the corecursive behavior of the registered
operations 
by describing how they produce one layer of observable data.
The definition in Figure~\ref{fig-init} depicts this for $\ctor$
and instantiates to the schematic behavior of $\SCons$ presented at the end of Example~\ref{exa-state}.

\begin{theorem} \label{th-init}\rm
$\initState$ is a well-formed corecursion state---i.e., it satisfies parametricity and well-behavedness.
\end{theorem}

\subsection{Advancing the Corecursion State}
\label{sec-adv}

{\looseness=-1 
The role of a corecursion state $(\ov{\K},\ov{f},\l)$ for $\J$ is to provide infrastructure for flexible corecursive definitions of functions $g$
between arbitrary sets $A$ and $\J$.
If nothing else is known about $A$, this is the end of the story.
However, assume that $\J$ is a component of $A$, in that $A$ is constructed from $\J$ (possibly along with other components).
For example, $A$ could be $\List\;\J$,
or $\J \times (\Nat \ra \List\;\J)$.
We capture this abstractly by assuming $A = \K\;\J$ for some functor $\K$.

}

In this case,
we have a fruitful situation of which we can profit for improving the corecursion state, and hence improving
the flexibility of future corecursive definitions. Under some
uniformity assumptions, $g$ itself can be registered as well behaved.

More precisely, assume that $g : \K\;\J \ra \J$ is defined by $g = \corecU\;s$ and that $s$ can be proved to be uniform in the following sense:
There exists a parametric function $\rho : \prod_{A \in \Set} \K\,(A \times \F\;A) \ra \F\;(\SS^*\,(\K\;A))$
such that $s = \rho \mathrel\circ \K \langle\id, \dtor\rangle$ (Figure~\ref{fig-integg}).
Then we can integrate $g$ as a well-defined operation as follows.
We define
$\nextState_{g}(\ov{\K},\ov{f},\l)$, the ``next'' corecursion state triggered by $g$,
as $(\ov{\K'},\ov{f'},\l')$, where
\begin{itemize}
\item $\ov{\K'} = (\K_1,\ldots,\K_n,\K)$ (similarly to $\SS$ versus $\ov{\K}$, we write $\SS'$ for the signature functor of $\K'$; note that
we essentially have $\SS' = \SS + \K$);
\item $\ov{f'} = (f_1,\ldots,f_n,g)$;
\item $\l' : \prod_{A \in \Set} \SS'\,(A \times \F\;A) \ra \F\,({\SS'}^*A)$ is
  defined as
$[\l \mathrel\circ \F\;\embL,\allowbreak\, \rho \mathrel\circ \F\;\embR]$ where $[\_,\_]$
is the case operator on sums,
which builds a function $[u,v] : B + C \ra D$ from two functions $u : B \ra D$ and $v : C \ra D$, and
  $\embL : \SS^* A \ra {\SS'}^* A$, $\embR: \SS^*\,(\K\;A) \ra {\SS'}^* A$ are the natural
embeddings into ${\SS'}^* A$.
\end{itemize}

\begin{theorem}\label{th-step}\rm
If $(\ov{\K},\ov{f},\l)$ is a well-formed corecursion state,
so is $\nextState_g\;(\ov{\K},\ov{f},\l)$.
\end{theorem}

In summary, we have the following scenario triggering the state's advancement:
\begin{enumerate}
\item One defines a new operation $g = \corecU\;s$.
\item One shows that $s$ factors through a parametric function $\rho$ and
$\K\,\langle\id, \dtor\rangle$ (as in Figure~\ref{fig-integg});
in other words, one shows that $g$'s corecursive behavior $s$ decomposes into a one-step
destruction
of the arguments and a parametric transformation (which is independent of $\J$).
\item The corecursion state is updated by $\nextState_{g}$.
\end{enumerate}

\begin{eexample}\rm \label{exa-integrate}
The operations $\onetwos$, $\opls$, $\oprd$, and $\oexp$ from \S\ref{sec-uptoExa}
are covered by this scenario.  For example,
assume that $\SCons$ and $\opls$ are registered as well behaved
at the time of defining $\oprd$ (cf.\ Example \ref{exa-state}).
Then $\K = (\lambda A.\;A^2)$ and
${\oprd} = \corecU\;s$, where
{
$$s = (\lambda (\xs,\ys).~(\!\begin{aligned}[t]
  & \hd\;\xs \times \hd\;\ys, \\[-\jot]
  & \eta\;(\xs,\tl\;\ys) \,\bb{\opls}\, \eta\;(\tl\;\xs,\ys)))\end{aligned}$$
}
The function $s$ decomposes into $\rho \mathrel\circ \K\,\langle\id,\langle \hd,\tl \rangle\rangle$, where
$$\textstyle\rho : \prod_{A \in \Set} (A \times (\Nat \times A))^2 \ra \Nat \times \SS^*\,(A^2)$$
is defined by
$\rho\,((a,\negvthinspace (m,\negvthinspace a')), (b,\negvthinspace (n,\negvthinspace b'))) = (m \times n, (a,\negvthinspace b') \vthinspace\bb{\opls}\vthinspace (a',\negvthinspace b))$,
which is clearly parametric.
The act 
of determining $\rho$ from
$s$ and $\K\,\langle\id,\langle \hd,\tl \rangle\rangle$
is syntax-directed.
\end{eexample}


\subsection{Coinduction Up-to}
\label{sec-coind}

In a proof assistant, specification mechanisms are not very useful unless
they are complemented by suitable reasoning infrastructure.
The natural counterpart of corecursion up-to is coinduction up-to.
In our incremental framework, the expressiveness of coinduction up-to grows
together with that of corecursion up-to.

We start with structural coinduction \cite{rutten00}, allowing to prove two elements of $\J$ equal by exhibiting an $\F$-bisimulation,
i.e., a binary relation $R$ on $\J$ such that whenever two elements $j_1$ and $j_2$ are related, their
$\dtor$-unfoldings are componentwise related by $R$.
\begin{center}
\AXC{$R\;j_1\,j_2$}
\AXC{$\forall j_1\,j_2 \in \J.\,R\;j_1\,j_2 \longrightarrow \Frel\;R\;(\dtor\;j_1)\;(\dtor\;j_2)$}
\BIC{$j_1 = j_2$}
\DP
\end{center}
Recall that our type constructors are not only functors but also relators.
The notion of ``componentwise relationship'' refers to $\F$'s relator structure $\Frel$.

Upon integrating a new operation $g$ (Section~\ref{sec-adv}), the coinduction rule is made more flexible
by allowing the $\dtor$-unfoldings to be componentwise related not only by $R$ but more generally
by a closure of $R$ that takes $g$ into account.

For a corecursion state $(\ov{\K},\ov{f},\l)$ and a relation $R : \J \ra \J \ra \Bool$,
we define $\CONG_{\ov{f}}\;R$, the $\ov{f}$-congruence closure of $R$, as the smallest equivalence relation
that includes $R$ and is compatible with each $f_i : \K_i\,\J \ra \J$:
$$
\forall z_1,z_2 \in \K_i\,\J.\,\Krel_i\,R\,z_1\,z_2 \lra \CONG_{\ov{f}}\;R\,(f_i\,z_1)\,(f_i\,z_2)
$$
where $\Krel_i$ is the relator associated with $\K_i$.

The next theorem supplies the reasoning counterpart of the definition principle stated in
Theorem~\ref{corec-char}. It can be inferred from recent, more abstract results \cite{rot-uptoCoind}.

\begin{theorem}\rm \label{th-coindUpTo}
The following coinduction rule up to $\ov{f}$ holds in the corecursion state $(\ov{\K},\ov{f},\l)$:
\begin{center}
\AXC{$R\;j_1\,j_2$}
\AXC{$\forall j_1\,j_2 \in \J.\,R\;j_1\,j_2 \longrightarrow \Frel\,(\CONG_{\ov{f}}\;R)\,(\dtor\;j_1)\;(\dtor\;j_2)$}
\BIC{$j_1 = j_2$}
\DP
\end{center}
\end{theorem}
Coinduction up to $\ov{f}$ is the ideal abstraction for proving equalities involving functions defined by corecursion up to $\ov{f}$:
For example, a proof of commutativity for $\oprd$ 
naturally relies on contexts involving $\opls$, because
$\oprd$'s corecursive behavior (i.e., $\oprd$'s $\dtor$-unfolding) depends on $\opls$.

\section{Mixed Fixpoints}
 \label{sec-mixed}




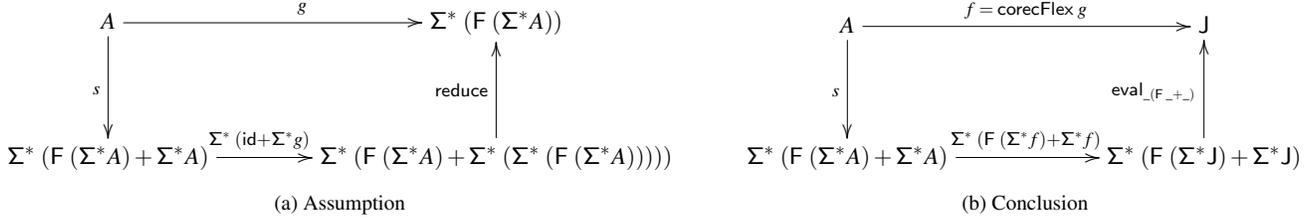
\begin{figure*}
        \centering
        \begin{subfigure}[b]{0.52\textwidth}
\centering
$\xymatrix@C=3pc@R=3pc{
 A
\ar^{g}[r]
\ar_{s}[d]    &  \SS^*\,(\F\,(\SS^*A))
                   \\
\SS^*\,(\F\,(\SS^*A) + \SS^*A)
\ar^{\SS^*\,(\id + \SS^*g)\hspace*{8ex}}[r]  & \SS^*\,(\F\,(\SS^*A) + \SS^*\,(\SS^*\,(\F\,(\SS^*A)))))  \ar^{\reduce}[u]
}$
                \caption{Assumption}
                \label{fig-mixed-asm}
        \end{subfigure}%
\enskip\quad
\vspace*{1ex}
\hspace*{1ex}
                \begin{subfigure}[b]{0.43\textwidth}
$\xymatrix@C=4.5pc@R=3pc{
 A
\ar^{f \;=\; \corecUU\;g}[r]
\ar_{s}[d]    &  \J
                   \\
\SS^*\,(\F\,(\SS^*A) + \SS^*A)
\ar^{\SS^*\,(\F\,(\SS^*f) + \SS^*f)}[r]  &  \SS^*\,(\F\,(\SS^*\J) + \SS^*\J) \ar^{\eval_{\_(\F\,\_ + \_)}}[u]
}$
                \caption{Conclusion}
                \label{fig-mixed-concl}
        \end{subfigure}%
\vspace*{-2ex}
\caption{Mixed fixpoint}
\label{fig-mixed}
\end{figure*}


When we write fixpoint equations to define a function $f$, we often want to
distinguish corecursive calls from calls that are sound for other reasons---for
example, if they terminate.
We model this situation abstractly
by a function $s : A \ra \SS^*\,(\F\,(\SS^*A) + \SS^*A)$. As usual for each $a$,
the shape of $s\;a$ represents the calling context for $f\;a$, with the occurrences of
the content items $a'$ in $s\;a$ representing calls to $f\;a'$. The new twist is that we now distinguish
guarded calls (captured by the left-hand side of $+$) from possibly unguarded ones
(the right-hand side of $+$).
%


We want to define a function $f$ with the behavior indicated by $s$, i.e., making the diagram in Figure~\ref{fig-mixed-concl} commute.
In the figure,
$+$ denotes the map function
$u + v : B + C \ra D + E$ built from two functions $u : B \ra D$ and $v : C \ra E$.
In the absence of pervasive guards, we cannot employ the corecursors directly to
define $f$.
However, if we can show that the noncorecursive calls eventually lead to a corecursive call, we will be able to employ $\corecUU$.
This precondition can be expressed in terms of a fixpoint equation.
According to Figure~\ref{fig-mixed-asm},
the call to $g$ (shown on the base arrow) happens only on the right-hand side of $+$, meaning that the intended
corecursive calls are ignored when ``computing'' the fixpoint $g$.
Our goal is to show that the remaining calls behave properly.

The functions $\reduce$ and $\eval$ that complete the diagrams of Figure~\ref{fig-mixed}
are the expected ones:
\begin{itemize}
\item The elements of $\SS^*\,(\F\,(\SS^*A))$ are formal-expression trees guarded on every path to the leaves,
and so are the elements $\SS^*\,(\F\,(\SS^*A) + \SS^*\,(\SS^*\,(\F\,(\SS^*A))))$, but with a more restricted shape;
$\reduce$ embeds the latter in the former: 
$\reduce = \mu \mathrel\circ \SS^*[\eta,\mu]$, where $\mu : \prod_{A \in \Set} \SS^*\,(\SS^*A) \ra A$ is the standard join
operation of the $\SS^*$-monad.
\item $\smash{\eval_{\_(\F\,\_ + \_)}}$ evaluates all the formal operations of $\SS^*$:
$$\smash{\eval_{\_(\F\,\_ + \_)}} = \eval \mathrel\circ \SS^*\,([\ctor \mathrel\circ \F\;\eval,\eval])$$
\end{itemize}

\begin{theorem}\rm \label{thm-mixed}
If there exists (a unique) $g : A \ra \SS^*\,(\F\,(\SS^*A))$ such that the diagram in Figure~\ref{fig-mixed-asm} commutes,
there exists (a unique) $f : A \ra \J$ such that the diagram in Figure~\ref{fig-mixed-concl} commutes, namely,
$\corecUU\;g$.
\end{theorem}

The theorem certifies the following procedure for making sense of a mixed fixpoint definition of a function $f$:
\begin{enumerate}
\item Separate the guarded and the unguarded calls (as shown in the codomain $\SS^*\,(\F\,(\SS^*A) + \SS^*A)$ of $s$).
\item Prove that the unguarded calls eventually terminate or lead to guarded calls (as witnessed by $g$).
\item Pass the unfolded guarded calls to the corecursor---i.e., take $f = \corecUU\;g$.
\end{enumerate}

\begin{eexample}\rm \label{exa-mixed}
The above procedure can be applied to define $\facC$,
$\primes : \Nat \ra \Nat \ra \Stream$, and $\catalan : \Nat \ra \Stream$,
while avoiding the unsound $\evil$ (\S\ref{sec-mix-exa}).
A simple analysis reveals that
the first self-call to $\primes$ is guarded while the second is not.
We define
$g : \Nat \times \Nat \ra \SS^*\,(\Nat \times \SS^*\,(\Nat \times \Nat))$ 
by
\begin{quote}
$g\;(m,n)  =  \keyw{if}~(m = 0 \mathrel\land n > 1) \mathrel\lor \GCD\;m\;n = 1\\
\phantom{g\;(m,n)  = {}} \keyw{then}~\eta\;(n, \eta\;(m \times n, n+1))\\
\phantom{g\;(m,n)  =  {}} \keyw{else}~g\;(m,n+1)$
\end{quote}
In essence, $g$ behaves like (the intended) $f$ except that the guarded calls are left
symbolic, whereas the unguarded calls are interpreted as actual calls to $g$.
One can show that $g$ is well defined by a standard termination argument.
This characteristic equation of $g$
is 
the commutativity of the diagram
determined by $s$ as in Figure~\ref{fig-mixed-asm},
where $s :  \Nat \times \Nat \ra \SS^*\,(\Nat \times \SS^*\,(\Nat \times \Nat) \mathrel+ \SS^*\,(\Nat \times \Nat))$ is
defined as follows
(with $\Inl$ and $\Inr$ being the left and right sum embeddings):
\begin{quote}
$s\;(m,n)  =  \keyw{if}~(m = 0 \mathrel\land n > 1) \mathrel\lor \GCD\;m\;n = 1\\
\phantom{s\;(m,n)  =  {}} \keyw{then}~\eta\;(\Inl\,(\eta\;(n, \eta\;(m \times n, n+1))))\\
\phantom{s\;(m,n)  =  {}} \keyw{else}~\eta\;(\Inr\,(\eta\;(m,n+1)))$
\end{quote}
Setting $\primes = \corecUU\;g$ yields the desired characteristic equation for
$\primes$
after simplification (cf.\ Example \ref{exa-state}).
\end{eexample}

The $\primes$ example has all unguarded calls in tail form, which makes the associated function $g$
tail-recursive. This need not be the case, as shown by the $\catalan$ example,
whose unguarded calls occur under
the well-behaved operation $\opls$. However, we do require that the unguarded calls occur in contexts formed by
well-behaved operations alone. 
After unfolding all the unguarded calls,
the resulting context that is to be handled corecursively
must be well behaved---this precludes unsound definitions like
$\evil$. 

\section{Formalization and Implementation}
\label{sec-for}

We formalized in Isabelle/HOL the metatheory of Sections~\ref{sec-meta} and \ref{sec-mixed}.
Essentially, this means that the results have been proved
in higher-order logic
with Infinity, Choice, 
and a mechanism for defining
types by exhibiting non-empty subsets of existing types. The logic is
comparable to Zermelo set theory with Choice (ZC) but weaker than ZFC.
The development would work for any class of functors that are relators
(or closed under weak pullbacks), contain basic functors (identity,
(co)products, etc.) and are closed under intersection, composition, and
have initial algebras and final coalgebra that can be represented in higher-order logic.
However, our Isabelle development focuses on a specific class \cite{traytel-et-al-2012}.


The formalization consists of two parts: The \emph{base} derives a corecursor
up-to from a primitive corecursor; the \emph{step} starts with a corecursor
up-to and integrates an additional well-behaved operation.

The base part starts by axiomatizing a functor $\F$
and defines a codatatype with nesting through $\F$:\,
  \keyw{codatatype} \,$\J$ $=$ $\ctor\;(\F\;\J)$.
(In~general, $\J$ could depend on type variables, 
but this is an orthogonal concern that would only clutter the formalization.)
Then the formalization defines the free algebra over $\F$ and
the basic corecursor seed $\l$ for initializing the state with
$\ctor$ as well behaved (Section~\ref{sec-init}). It also needs to lift
$\l$ to the free algebra, a technicality that was omitted in the
presentation. Then it defines $\eval$ and other necessary structure
(Section \ref{sec-state}). Finally, it introduces $\corecU$ and $\corecUU$
(Section~\ref{sec-corec-princ}) and derives the corresponding
coinduction principle (Section~\ref{sec-coind}).

From a high-level point of view, the step part has a somewhat similar structure to the base. It
axiomatizes a domain functor $\K$ and a parametric function $\rho$ associated
with the new well-behaved operation $g$ to integrate.
Then it extends the signature to include $\K$, defines the
extended corecursor seed $\l'$, and lifts $\l'$ to the free algebra.
Next, it defines the parameterized $\eval_g$ and other infrastructure
(Section~\ref{sec-adv}). Finally, it introduces $\corecU$ and $\corecUU$
for the new state and derives the coinduction principle.

The process of instantiating the metatheory to particular user-specified codatatypes
is automated by a prototype tool: the user points to a particular codatatype (typically defined using
Isabelle's existing (co)datatype specification language \cite{blanchette-et-al-2014-impl}), and then the tool
takes over and instantiates the generic corecursor to the indicated type, provinding the concrete corecursion
and mixed recursion-corecursion theorems.
The stream and
tree examples presented in Section~\ref{sec-exa} 
%
have all been obtained with this tool.
As a larger case study, we formalized all the examples
from the extended version of Hinze and James's study \cite{hinze-adventure}. 
The parametricity proof obligations were discharged by Isabelle's
parametricity prover \cite{huffman-lifting}.
The mixed recursion--corecursion definitions were done using Isabelle's
facility for defining terminating recursive functions \cite{krauss-fun}.


Unlike Isabelle's {\relax primitive} (co)recursion mechanism \cite{blanchette-et-al-2014-impl},
our tool currently lacks syntactic sugar support,
so it still requires some boilerplate from the user, namely the explic invocation of the corecursor
and the parametricity prover: these are just a few extra lines of script per definition, and therefore the tool is
also usable in the current form.
Following the design of its primitive ancestor, 
its envisioned fully user-friendly extension will replace the explicit invocation of the
corecursor with a \keyw{corec} command,
allowing users to specify a function $\const{f}$ corecursively and
then performing the following steps (cf.\ Example~\ref{exa-defUpToInst}):
\begin{enumerate}
\item Parse the specification of $\const{f}$ and synthesize arguments to the
current, most powerful corecursor.
\item Define $\const{f}$ in terms of the corecursor.
\item Derive the original specification from the corecursor theorems.
\end{enumerate}
Passing the \keyw{well\_behaved} option to \keyw{corec}
will additionally invoke the following procedure (cf.\ Example~\ref{exa-integrate}):
\begin{enumerate}
\item[4.] Extract a polymorphic function $\rho$ from the specification of $\const{f}$.
\item[5.] Automatically prove $\rho$ parametric or pass the proof obligation to the user.
\item[6.] Derive the new strengthened corecursor and its new coinduction principle.
\end{enumerate}
The \keyw{corec} command will be complemented by an additional command,
tentatively called \keyw{well\_behaved\_for\_corec}, for registering arbitrary operations
$\const{f}$ (not necessarily defined using
\keyw{corec}) as well behaved. The command will ask the user to provide a
corecursive specification of $\const{f}$ as a lemma of the form
$\const{f}\;\ov{x} = \Cons\,\ldots$ \,and then perform steps 4~to~6.
%
%
%
The \keyw{corec} command will become
stronger and stronger as more well-behaved operations are registered.

The following Isabelle theory fragment gives a flavor of the envisioned functionality
from the user's point of view:

\begin{quote}
  \keyw{codatatype} \,$\Stream\; A$ $=$ $\SCons\;(\hd{:}\; A)\;(\tl{:}\; \Stream\; A)$
\\[2\jot]
  \keyw{corec} (\keyw{well\_behaved}) ${\opls} : {}\Stream \ra \Stream \ra \Stream$ 
  \\
  \hbox{}\quad$\xs \opls \ys  =  {\SCons}\;(\hd\;\xs + \hd\;\ys)\;(\tl\;\xs \opls \tl\;\ys)$
\\[2\jot]
  \keyw{corec} (\keyw{well\_behaved}) ${\oprd} : \Stream \ra \Stream \ra \Stream$ 
  \\
  \hbox{}\quad$\xs \oprd \ys  =  {\SCons}\;(\hd\;\xs \times \hd\;\ys)\;\\
  \phantom{\hbox{}\quad\xs \oprd \ys  =  {\SCons}\;}((\xs \oprd \tl\;\ys) \opls (\tl\;\xs \oprd ys))$
\\[2\jot]
  \keyw{lemma} \textit{$\opls$\_commute}:\enskip $\xs \opls \ys = \ys \opls \xs$ \\
  \hbox{}\quad\keyw{by} (\textit{coinduction arbitrary}:\ \textit{xs ys rule}:\ \textit{stream.coinduct})~\textit{auto}\kern-200mm
\\[2\jot]
  \keyw{lemma} \textit{$\oprd$\_commute}:\enskip $\xs \oprd \ys = \ys \oprd \xs$ \\
  \keyw{proof} (\textit{coinduction arbitrary}:\ \textit{xs ys rule}:\ \textit{stream.coinduct\_upto})\kern-200mm \\
  \hbox{}\quad\keyw{case} \textit{Eq\_stream} \\
  \hbox{}\quad\keyw{thus} \textit{?case} \keyw{unfolding} \textit{tail\_$\oprd$} \\
  \hbox{}\qquad\keyw{by} (\textit{subst} \textit{$\opls$\_commute}) (\textit{auto intro}: \textit{stream.cl\_$\opls$}) \\
  \keyw{qed}
\end{quote}



\section{Related Work}
\label{sec-rel}

There is a lot of relevant work, concerning both the metatheory and applications
in proof assistants and similar systems. We referenced some of the most closely
related work in the earlier sections. Here is an attempt at a more systematic overview.

\paragraph{Category Theory.}
The notions of corecursion and coinduction up-to started with process algebra
\cite{san-bis, RuttenProcAsTerms} before they were
recast in the abstract language of category theory
\cite{BartelsGeneralizedCoind,turi-plotkin97,klin11-bialgebras,milius-modular,jacobs06-distrib,rot-uptoCoind,hinze-adventure}.
%
Our approach owes a lot to this theoretical work, and indeed formalizes some state-of-the-art
category theoretical results on corecursion and coinduction up-to \cite{milius-modular,rot-uptoCoind}.
Besides adapting existing results to higher-order logic 
within an incremental corecursor cycle,
we have also extended the state of the art with a sound mechanism for mixing recursion with corecursion up-to.


Category theory provides an impressive body of {\relax abstract} results
that can be applied to solve concrete problems elegantly.
Proof assistants have a lot to benefit from category theory, as we hope to have demonstrated
with this paper.
There has been prior work on integrating coinduction up-to techniques from
category theory into these tools. Hensel and Jacobs
\cite{hensel-interatedRecursion} illustrated the categorical approach to
(co)data\-types in PVS via axiomatic declarations of various flavors of
trees with (co)recursors and proof principles. Popescu and Gunter proposed
incremental coinduction for a deeply embedded proof system in Isabelle\slash HOL
\cite{pop-Coind}. Hur et al.\ \cite{HurNDV13} extended Winskel's
\cite{winskel-nu} and Moss's \cite{moss-param} parameterized coinduction and
studied applications to Agda, Coq, and Isabelle\slash HOL. Endrullis et al.\
\cite{endrulis-circ} developed a method to perform 
up-to coinduction in Coq adapting insight from behavioral logic \cite{circCALCO09theory}.
To our knowledge, no prior work has realized corecursion up-to in a
proof assistant.

\paragraph{Ordered Structures and Convergence.}
A number of approaches to define functions on infinite types are based on domain
theory, or more generally on ordered structures and notions of convergence, including Matthews
\cite{matthews-rec-coind}, Di~Gianantonio and Miculan \cite{miculan-unifying},
Huffman \cite{huffman-2009}, and Lochbihler and H\"olzl
\cite{lochbihler-hoelzl-2014}. These are not directly comparable to our work
because they do not guarantee productivity or otherwise offer total programming.
They also force the user to switch to a different, richer universe of domains or
to define ordered structures and perform continuity proofs (although Matthews
shows that this process can be partly automated).

Strictly speaking, our approach does not guarantee productivity either. This is an
inherent limitation of the semantic
(shallow embedded)
approach in HOL systems, which do not specify a computational model (unlike Agda and Coq).
Productivity can be argued informally by inspecting the characteristic corecursion equations.

\paragraph{Syntactic Criteria.}
Proof assistants based on type theory include checkers for termination of
recursion functions and productivity of corecursive functions. These checkers
are part of the system's trusted code base; bugs can lead to inconsistencies, as
we saw for Agda \cite{traytel-2014-agda} and Coq \cite{denes-2013-coqml}.%
\footnote{In all fairness, we should mention that critical bugs were also
found in the primitive definitional mechanism of our proof assistant of
choice, Isabelle \cite{kuncar-2015-cpp}. Our point is not that brand B is superior to brand A, but
rather that it is generally desirable to minimize the amount of trusted code.}
\,For users, such syntactic criteria are also inflexible; for example, Coq allows
more than one constructor to appear as guards but is otherwise limited to
primitive corecursion.

To the best of our knowledge, the only deployed system that explicitly supports mixed
recursive--corecursive definitions 
is Dafny. Leino and
Moskal's paper \cite{leino-moskal-2014} triggered our interest in the topic.
Unfortunately, the paper is not entirely clear about the supported fragment. A naive
reading suggests that the inconsistent $\evil$ example from
Section~\ref{sec-mix-exa} is allowed, as was the case with earlier versions of
Dafny. Newer versions reject not only $\evil$ but also the legitimate
$\catalan$ function from the same subsection.

\paragraph{Type Systems.}
A more flexible alternative to syntactic criteria is to have users annotate
the functions' types with information that controls termination and
productivity. Approaches in these category include
fair reactive programming \cite{fra1,usrp1,cave-et-al-2014},
clock variables \cite{mcbride-productive,clouston-et-al-2015}, and
sized types \cite{abel-2004}. Size types are implemented in
MiniAgda \cite{abel-2010-miniagda}
and in newer versions of Agda, in conjunction with a destructor-oriented
(copattern) syntax for corecursion \cite{
abelP-2013}.
These approaches, often featuring a blend of type systems and notions of convergence, achieve a
higher modularity and trustworthiness, by moving away from purely syntactic
criteria and toward semantic properties.
By carefully tracking sizes and timers, they allow
for more general contexts than our well-behavedness criterion.
Our approach captures a 1--1 contract: A well-behaved
function can destroy one constructor to produce one. A function $\const{f}$ that would,
map the stream $a_1, a_2, \ldots$ to $a_1, a_1, a_2, a_2, \ldots$ would have a 1--2
contract. And a function $\const{g}$ mapping $a_1, a_2, a_3, a_4, \ldots$ to
$a_1 + a_2, a_3 + a_4, \ldots$ would require a 2--1 contract. The composition
$\const{g} \mathrel\circ \const{f}$ would yield a 1--1 contract and could in
principle appear in a corecursive call context, but our framework does not allow it.

Clock variables and sized types require an extension to the type system and
burden the types. These general contracts must be specified by the user and
complicate the up-to corecursion principle; the arithmetic that ensures that
contracts fit together would have to be captured in the principle, giving rise
to new proof obligations. In contrast, well-behaved functions can be freely
combined. This is the main reason why we can claim it is a ``sweet spot.''

There is a prospect of embedding our lighter approach into such heavier but more precise frameworks.
Our well-behaved
operators possibly form the maximal class of context functions
requiring no annotations (in general), amounting to a lightweight subsystem
of Krishnaswami and Benton's type system \cite{usrp1}.


\section{Conclusion}
\label{sec-conc}

We presented a formalized framework for deriving rich corecursors that can be used to
define total functions producing codatatypes. The corecursors gain in
expressiveness with each new corecursive function definition that satisfies a
semantic criterion. They constitute a significant improvement over the state of
the art in the world of proof assistants based on higher-order logic,
including HOL4, HOL Light, Isabelle\slash HOL, and PVS.
Trustworthiness is attained at the cost of elaborate 
constructions.
Coinduction being somewhat counterintuitive,
we argue that these safeguards are well worth the effort.
As future work, we want
to transform our prototype tool into a solid implementation inside
Isabelle\slash HOL. 


Although we emphasized the foundational nature of the framework, many of the
ideas equally apply to systems with built-in codatatypes and
corecursion. 
One could imagine extending the productivity check of
Coq to allow corecursion under well-behaved operations, linking a syntactic
criterion to a semantic property, as a lightweight alternative to
clock variables and sized types. The emerging infrastructure
for parametricity in Coq \cite{bernardy-et-al-2012-param,keller-lasson-2012}
would likely be a useful building block.




\paragraph{Acknowledgment.}
Tobias Nipkow made this work possible.
Stefan Milius guided us through his highly relevant work on abstract GSOS rules.
Andreas Abel and Rustan Leino discussed their tools and shared
their paper drafts and examples with us.
Mark Summer\-field suggested many textual improvements.
Reviewers 
provided useful comments and indicated related work.
Blanchette was supported by the Deutsche
Forschungs\-gemein\-schaft (DFG) project \relax{Hardening the Hammer} (grant
Ni\,491\slash 14-1).
Popescu was supported by the DFG project
\relax{Security Type Systems and Deduction} (grant Ni\allowbreak\,491\slash 13-2) as part
of the program \relax{Reliably Secure Software Systems} (RS\textsuperscript{3},
priority program 1496).
Traytel was supported by the DFG program \relax{Program and Model Analysis}
(PUMA, doctorate program 1480).
 The authors are listed alphabetically.

\begin{raggedright}
\small
\bibliography{bib}
\end{raggedright}


\end{document}

%% file: defs.tex
\usepackage{amssymb}

\newcommand{\leftOut}[1]{}
\newcommand{\comment}[1]{}

\newbox\boxA

\renewcommand{\epsilon}{\varepsilon}

\newcommand\vvthinspace{\kern.041667ex}
\newcommand\vthinspace{\kern.08333ex}
\newcommand\negvthinspace{\kern-.08333ex}
\newcommand\negvvthinspace{\kern-.041667ex}

\newcommand\TC{\mathsf}
\newcommand\keyw[1]{\texttt{#1}}

\newcommand\CHOPFROMUN{.25}
\newcommand\UN{{\setbox\boxA=\hbox{\_}\usebox\boxA\kern-\CHOPFROMUN\wd\boxA{\color{white}\vrule height 0ex depth .444ex width \CHOPFROMUN\wd\boxA}\kern-\CHOPFROMUN\wd\boxA}}

\newtheorem{theorem}{Theorem}
\newtheorem{prop}[theorem]{Proposition}
\newtheorem{eexample}[theorem]{Example}
\newenvironment{proofx}{
  \upshape\noindent{\it Proof.}\enskip\ignorespaces}{\qed\vspace{\topsep}}

\newcommand{\ov}{\overline}

\renewcommand{\phi}{\varphi}

\renewcommand{\P}{\TC{P}}

\renewcommand{\iff}{\allowbreak\mathrel{\leftarrow\nobreak\kern-1.6ex\rightarrow}\allowbreak} 
\newcommand{\iffla}{\allowbreak\mathrel{\leftarrow\nobreak\kern-1.55ex-}\allowbreak}

\newcommand{\ra}{\rightarrow}
\newcommand{\Ra}{\Rightarrow}
\newcommand{\lra}{\longrightarrow}

\newcommand{\llam}{{\llam}}

\renewcommand{\l}{\Lambda}
\newcommand{\talg}{{{\mathsf{op}}}} 
\renewcommand{\eta}{{{\mathsf{leaf}}}}
\renewcommand{\mu}{{{\mathsf{flat}}}}
\newcommand{\ctor}{{{\mathsf{ctor}}}}
\newcommand{\dtor}{{{\mathsf{dtor}}}}
\newcommand{\lfp}{{{\mathsf{lfp}}}}
\newcommand{\gfp}{{{\mathsf{gfp}}}}
\newcommand{\inc}{{{\mathsf{inc}}}}
\newcommand{\evil}{{{\mathsf{nasty}}}}

\newcommand{\Grel}{{{\mathsf{Grel}}}}

\newcommand{\initState}{{{\mathsf{init\hspace*{-0.075ex}State}}}}
\newcommand{\nextState}{{{\mathsf{next\hspace*{-0.075ex}State}}}}
\newcommand{\embL}{{{\mathsf{emb\hspace*{-0.075ex}L}}}}
\newcommand{\embR}{{{\mathsf{emb\hspace*{-0.075ex}R}}}}

\newcommand{\Frel}{{{\mathsf{Frel}}}}
\newcommand{\Krel}{{{\mathsf{Krel}}}}

\newcommand{\onetwos}{{{\mathsf{onetwos}}}}

\newcommand{\reduce}{{{\mathsf{reduce}}}}

\newcommand{\Call}{{{\eta}}}

\newcommand{\eval}{{{\mathsf{eval}}}}
\newcommand{\corec}{{{\mathsf{corecPrim}}}}
\newcommand\corecU{\mathsf{corecTop}}
\newcommand\corecUU{\mathsf{corecFlex}}
\newcommand{\mapS}{{{\mathsf{smap}}}}
\newcommand{\listrel}{{{\mathsf{rel}}}}

\newcommand{\Children}{{{\mathsf{sub}}}} 
\newcommand{\Label}{{{\mathsf{val}}}} 

\newcommand{\hd}{{{\mathsf{head}}}}
\newcommand{\tl}{{{\mathsf{tail}}}}
\newcommand{\SCons}{{{\mathsf{S\hspace*{-0.2ex}Cons}}}}
\newcommand{\Cons}{{{\mathsf{Cons}}}}

\newcommand{\Node}{{{\mathsf{Node}}}}
\newcommand{\opls}{\mathrel\oplus}
\newcommand{\oprd}{\mathrel\otimes}
\newcommand{\oexp}{{{\mathsf{exp}}}}
\newcommand{\primes}{{{\mathsf{primes}}}}
\newcommand{\GCD}{{{\mathsf{gcd}}}}
\newcommand{\everyOther}{{{\mathsf{everyOther}}}}
\newcommand{\fibA}{{{\mathsf{fibA}}}}
\newcommand{\fibB}{{{\mathsf{fibB}}}}
\newcommand{\facA}{{{\mathsf{facA}}}}
\newcommand{\facB}{{{\mathsf{facB}}}}
\newcommand{\facC}{{{\mathsf{facC}}}}

\newcommand{\fac}{{{\mathsf{fac}}}}
\newcommand{\SUP}{{{\mathsf{sup}}}}
\newcommand{\catalan}{{{\mathsf{cat}}}}
\newcommand{\CONG}{{{\mathsf{cl}}}}
\newcommand{\GUARD}[1]{{{\underline{#1}}}}

\newcommand{\id}{\mathsf{{id}}}

\newcommand{\F}{{\TC{F}}}

\newcommand{\J}{{\TC{J}}}
\renewcommand{\SS}{{\TC{\Sigma}}}
\newcommand{\K}{{\TC{K}}}

\newcommand{\G}{{\TC{G}}}

\newcommand{\map}{\mathsf{{map}}}
\newcommand{\zip}{\mathsf{{zip}}}

\newcommand{\fimage}{\mathsf{{fimage}}}

\newcommand{\Fmap}{\mathsf{{Fmap}}}

\newcommand{\Inl}{\mathsf{{Inl}}}
\newcommand{\Inr}{\mathsf{{Inr}}}

\newcommand{\snd}{\mathsf{{snd}}}

\newcommand{\xs}{{\mathit{xs}}}
\newcommand{\ys}{{\mathit{ys}}}

\newcommand\Set{\TC{Set}}

\newcommand\Stream{\TC{Stream}}
\newcommand\Nat{{\TC{Nat}}}

\newcommand{\Bool}{{\TC{Bool}}}
\newcommand{\Unit}{{\TC{Unit}}}

\newcommand{\List}{{\TC{List}}}

\newcommand{\Tree}{{\TC{Tree}}}

\newcommand{\FSet}{{\TC{FinSet}}}

\newmuskip\originalthinmuskip
\originalthinmuskip=\thinmuskip
\newmuskip\tinyGapMu
\renewcommand\ldots{\mathinner{.\mskip\originalthinmuskip .\mskip\originalthinmuskip .}}
\renewcommand\cdots{\mathinner{{\cdot}\mskip\originalthinmuskip {\cdot}\mskip\originalthinmuskip {\cdot}}}

\newcommand\bb[1]{\fbox{$#1$}}

\newcommand\const[1]{{\ensuremath{\TC{#1}}}}
\newcommand\PARA[1]{\subsubsection*{#1}}
\newcommand\FUNCTOR[2]{\PARA{#1~\ensuremath{#2}\kern.05ex}} 

\newcommand\Smash[1]{\kern-200mm\smash{#1}\kern-200mm}
\newcommand\SubItem[2]{\indent\hbox to \leftmargini{\hfill#1\enskip}#2}

\newcommand\XDot{\raise1ex\hbox{\Large.\kern.1em}}

